\documentclass[letterpaper]{JHEP}
\usepackage{epsfig,amssymb}

\newcommand{\wt}{\widetilde }
\newcommand{\p}{\partial}
\newcommand{\pu}{\partial^}
\newcommand{\pd}{\partial_}
\newcommand{\beq}{\begin{equation}}
\newcommand{\eeq}{\end{equation}}
\newcommand{\br}{\begin{eqnarray}}
\newcommand{\er}{\end{eqnarray}}
\newcommand{\brs}{\begin{eqnarray*}}
\newcommand{\ers}{\end{eqnarray*}}
\newcommand{\Z}{\mathbb{Z}}
\newcommand{\re}{{\rm Re}\,}
\newcommand{\im}{{\rm Im}\,}
\newcommand{\none}{${\cal N}${=}1 }
\newcommand{\nn}{${\cal N}${=}2 }

\title{BPS Equations, BPS States, and Central Charge 
	of \nn Supersymmetric Gauge Theories}
\author{Ioan A. Popescu$^a$ and Alfred D. Shapere$^{a,b}$\\

	$^a$Department of Physics, 
	University of Kentucky, 
        Lexington KY 40502\\[1mm]

	$^b$Center for Theoretical Physics, 
	M.I.T. 6-306, 
        Cambridge MA 02139\\[1mm]

        E-mail: \email{popescu@pa.uky.edu, shapere@pa.uky.edu}}

\abstract{We derive the central charge and BPS equations from the low-energy effective action for \nn  SU(2) Yang-Mills theory in the Coulomb phase, using a systematic, canonical procedure. We then obtain solutions for monopole and dyon BPS states, whose core structure is described by a dual Lagrangian containing the monopole or dyon as a fundamental field.  Spherically symmetric states possess a shell of charge at a characteristic radius.}
\preprint{ }
\begin{document}

\section{Introduction}

The remarkable solvability properties of \nn  gauge theories have led to much insight into the structure of four--dimensional field theories.  In particular, Seiberg and Witten's exact solution of the low--energy effective Lagrangian and BPS spectrum of the Coulomb phase of \nn  SU(2) Yang-Mills theory, and subsequent work, has provided us with solvable examples of gauge theories exhibiting many of the strong--coupling phenomena associated with QCD \cite{SW1,SW2}.

BPS states appear at several key steps in Seiberg and Witten's derivation, which leads to a formula for their exact mass spectrum. A crucial role is played by the central charge $Z$ of the \nn  algebra in the presence of a charged state, which appears in the BPS mass inequality
$$
M\ge \sqrt2 |Z|
$$
The masses of BPS states saturate this inequality, and are therefore determined by the central charge, which in turn depends on the vevs and masses that parameterize the moduli space of Coulomb vacua. As given by Seiberg and Witten for pure SU(2), the formula for the central charge is
\beq\label{Z}
Z=a_\infty n_e + a_{D\infty} n_m
\eeq
where $a_\infty$ is the asymptotic expectation value of the scalar component $a$ of the unbroken U(1) vector multiplet, and $a_{D\infty}$ is the asymptotic value of its dual, $a_D$, which is expressed in terms of the \nn  prepotential ${\cal F}(a)$ as $a_D\equiv \partial {\cal F}/\partial a$.

By following the dependence of $Z$ on $a_\infty$, one finds that the BPS spectrum changes discontinuously along codimension 1 surfaces of marginal stability, and that on  codimension 2 surfaces in the moduli space, certain states become massless.  Each massless surface is associated with a duality transformation --- the monodromy that results when one traverses a circuit enclosing the massless surface.  Together, these monodromies generate the duality group, which combined with holomorphicity and weak-coupling data, completely determines the \nn  prepotential.

In view of the crucial role played by BPS states in the Seiberg--Witten solution, a systematic derivation of the BPS equations based entirely on supersymmetry is desirable. 
Such an approach might be useful 
%This is especially desirable 
if one wishes to study the spectrum of stable BPS dyons, since
although dyons of all possible charges are conjectured to exist
\cite{Bergman}, only dyons with at most one unit of magnetic charge
can be supersymmetric, {\it i.e.}, BPS. One of the main aims of this
paper will be to give a rigorous derivation of the central charge formula and BPS equations. 

Another goal of this paper is to study properties of BPS soliton solutions within  the framework of the U(1) effective field theory.  Previous work along these lines has focused on solitons of the SU(2) effective theory, that is, with massive $W$ bosons included \cite{SW1,CRV}. Other authors have obtained 
useful information about soliton spectrum by embedding 
the gauge theory in string or M theory \cite{KLMVW, Sen,Faya,Bergman1, GHZ, BF1, MNS, DHIZ, BF2, SV}.
 However, since dyons can have masses below the $W$ mass scale $M_W$, one expects to find dyons in the soliton sector of even the fully truncated U(1) theory. This is the approach that has been taken more recently in \cite{RSVV, ArgyresNarayan, RV}. The U(1) effective Lagrangian provides a framework of minimal complexity and maximal solvability for studying the properties and dynamics of dyons, which has been used in these papers to study prong solutions from a purely field theoretic point of view. A further motivation for working in the context of the U(1) theory, which we will exploit, is that it has a dual description, valid even in the presence of massless dyons.  We shall find this dual theory essential for understanding the core structure of the dyon solutions. 

We begin in section 2 with a canonical derivation of the Noether supercharges.   The application of Noether's theorem is straightforward, if algebraically involved, and agrees with the result of \cite{Iorio2}. One interesting result we will obtain along the way is that the supersymmetry variation of the action itself is not generally zero in the presence of magnetic charges (although it does vanish if the charges are BPS). 

In section 3, we compute the Dirac bracket of the two supercharges in order to obtain a formula for the central charge.  
The central charge formula for the U(1) effective theory 
has been derived directly from the supersymmetry algebra \cite{Iorio1,Iorio2} only relatively recently, and as we shall explain, the derivation is incomplete. Specifically, 
the expression we will find for $Z$ differs from (\ref{Z})
by integral expressions that are  arbitrarily set to zero in 
\cite{Iorio1,Iorio2},
but which depend sensitively on the core boundary conditions
of the appropriately charged BPS solitons.\footnote{
These additional terms have been found independently
in \cite{ArgyresNarayan}.}
A major theme of this paper will be to 
explain why the additional terms are in fact equal to zero, leaving us with the standard formula (1.1). 

%This formula appears to differ from the Seiberg--Witten formula by integral expressions, which can be made to vanish by an assumption about the core structure of charged solitons. This assumption will be borne out when we construct BPS solitons explicitly. 

In Section 4, we obtain the BPS field equations, discuss their duality properties, and find monopole and dyon solutions. The singularities of these solutions at the core motivate us in Section 5 to study the dual Lagrangian valid in the vicinity of the point in moduli space where a monopole becomes massless. 
%We will study the core structure of BPS dyons by means of a dual Lagrangian which contains the 
%dyon as a local field,  
%and 
The Lagrangian contains the monopole as a local field. Its
BPS equations imply that the monopole charge
density can only be nonzero on a spherical shell with a 
calculable radius $r_0$. This is the radius at which   
the BPS dyon field's local mass, expressed as a linear 
combination of $a(r)$ and $a_D(r)$, goes to zero. 
For the magnetic monopole, the 
radius turns out to be inversely proportional to $|a_\infty -
a(r_0)|$. Thus, by taking $a_\infty$ close to $a(r_0)$, 
the size of the
sphere can be made as large as desired relative to the
short--distance cutoff, $\Lambda^{-1}$.
 
%Solving the BPS equations implied by this Lagrangian then yields a solution already suggested by the calculation of the central charge, with the magnetic charge 
%distributed on a sphere where the monopole field is effectively massless, and 
%carried by zero modes of the monopole field  localized on the sphere. 

In Section 6 we show that the solutions of Section 5 can be obtained from the original Lagrangian by including electric and magnetic source terms.  When these terms are included, the additional term in the supersymmetry variation that was found in Section 2 is canceled, and the Lagrangian becomes manifestly supersymmetric in the presence of magnetic charges. 

We close our introduction 
with the remark that similar structures to 
our shells of charge have appeared elsewhere: in 
the finite--energy BIon solutions of \cite{GKMTZ}
and as gravitational ``empty holes'' in \cite{Denef, DGR}. 
In fact, we believe that our solutions are nothing but empty 
holes in the gravitational decoupling limit. 
What distinguishes our approach is the use of the dual
Lagrangian to describe the vicinity of the spherical shell of 
dyon charge, which allows us to analyze this region at weak coupling.  
It will be interesting to apply this method to the
analysis of prong solutions, to improve our understanding of 
their core structure, and perhaps to obtain a field theoretic
derivation of their BPS spectrum. 

\bigskip

\section{U(1) Effective Lagrangian, Equations of Motion, and Supersymmetry Generators}

\bigskip
We begin by writing down the U(1) effective Lagrangian and equations of motion, from which we derive the canonical \nn  supersymmetry generators using Noether's theorem. We shall find that the supersymmetry variation of the Lagrangian contains an additional term which explicitly breaks some of the supersymmetry when magnetic monopoles are present.

The low-energy effective Lagrangian of the \nn  super Yang-Mills theory in the Coulomb phase, with gauge group SU(2) spontaneously broken down to U(1), has the following expression in terms of \none  superfields \cite{SW1}:
\beq
{\cal L}=\frac{1}{4\pi}\,{\rm Im}\left[\int d^2\theta\int d^2\bar\theta\,{\p{\cal F}\over\p\Phi}\,\bar\Phi\,+\,{1\over 2}\,\int d^2\theta\,{\pu 2{\cal F}\over\p\Phi^2}\,W^\alpha W_\alpha\right]
\eeq
As shown in Appendix B (where also our notations and conventions are introduced), the Lagrangian (1) has the following expression in terms of component fields
\br
{\cal L}&=&\frac{1}{4\pi}{\rm Im}\left\{\tau\left[\partial_\mu a\,\partial^\mu\bar a-i\psi\,\sigma^\mu\partial_\mu \bar\psi-i\lambda\,\sigma^\mu\partial_\mu \bar\lambda-\frac{1}{4}F_{\mu\nu}(F^{\mu\nu}\!-\!i\widetilde F^{\mu\nu})\!+\!f\bar f+\frac{1}{2}D^2\right]\right.\nonumber\\
&&+\,\frac{\tau'}{2}\,\left[i\sqrt{2}D(\psi\lambda)-\sqrt{2}(\lambda\,\sigma^{\mu\nu}\psi)F_{\mu\nu}-(\psi\psi)\bar f-(\lambda\lambda)f\right]
%\nonumber\\
%&&
+\,\left.\frac{\tau''}{4}\,(\psi\psi)(\lambda\lambda)\right\}\label{1}
\er
The auxiliary fields $f$ and $D$ can be eliminated using their constraint equations:
$$
f={1\over4i}\,{1\over {\rm Im}\,\tau}\left(\tau'\,\psi\psi-\bar\tau'\,\bar\lambda\bar\lambda\right),\quad D=-{1\over2\sqrt{2}}\,{1\over {\rm Im}\,\tau}\left(\tau'\,\psi\lambda+\bar\tau'\,\bar\psi\bar\lambda\right).
$$
The equation of motion for the gauge field $A_\nu$ is
\beq\label{eqgauge}
-\frac{1}{4\pi}\,\partial_\mu\,{\rm Im}\,\left[\tau(F^{\mu\nu}-i\widetilde F^{\mu\nu})\,+\,\sqrt{2}\,\tau'(\lambda\,\sigma^{\mu\nu}\,\psi)\right]=0,\quad\nu=0,1,2,3.
\eeq
The equation of motion obtained by varying the scalar field $\bar a$ is
$$
\pu\mu\left[2i({\rm Im}\,\tau)\pd\mu a\right]=-\bar\tau'\left[\pd\mu a\pu\mu\bar a-{1\over4}\,F_{\mu\nu}(F^{\mu\nu}+i\widetilde F^{\mu\nu})\right],
$$
where we have omitted the terms involving fermion fields\footnote{In this paper we will keep fermions in our discussion whenever necessary. However, the preceding equation will be applied only in a purely bosonic context.}. This last equation can also be written as
\beq\label{eqfora}
\partial_{\mu}\tau\partial^{\mu}a+2i({\rm Im}\,\tau)\partial_\mu\partial^\mu a-\frac{1}{4}\bar \tau'F_{\mu\nu}(F^{\mu\nu} + i\widetilde F^{\mu\nu})=0
\eeq

\bigskip
Using the variations of the component fields under an \none  transformation with parameter $\xi$, we eventually obtain (see Appendix D) the \none  supersymmetry variation of the Lagrangian,
\br\label{deltaL}
\delta {\cal L}&=&\partial_\mu\left\{\frac{1}{4\pi}\,{\rm Im}\,\left[\sqrt{2}\,\tau(\xi\psi)\,\partial^\mu\bar a-2\sqrt{2}\,\bar\tau(\xi\,\sigma^{\mu\nu}\psi)\,\partial_\nu\bar a
-i\sqrt{2}\,\bar\tau'(\bar\psi\bar\lambda)(\xi\,\sigma^\mu\bar\lambda)
\right.\right.\nonumber\\
&&+i\bar\tau(\xi\,\sigma_\nu\bar\lambda)(F^{\mu\nu}+i\widetilde F^{\mu\nu})-\left.\left.i\sqrt{2}\,\bar\tau f(\xi\,\sigma^\mu\bar\psi)+\bar\tau D(\xi\,\sigma^\mu\bar\lambda)
\right]\right\}\nonumber\\
&&+\frac{1}{4\pi}\,{\rm Im}\,[2\,\bar\tau(\xi\,\sigma^\nu\bar\lambda)(\partial^\mu\widetilde F_{\mu\nu})].
\er
The last term in (\ref{deltaL}) remains outside the total divergence\footnote{Such a term will always arise from Lagrangians containing an \none  vector superfields \cite{Wei}.}. In the absence of magnetic charges, this term vanishes by the Bianchi identity. However, in the present theory, magnetic monopoles do exist and we have, in terms of the magnetic number density $\nu_m$,
$$
\pd i\widetilde F^{i0}=4\pi \nu_m.
$$
Thus the last term in (\ref{deltaL}) cannot be ignored and appears to break supersymmetry. However, as we will show in sections 5 and 6, a consistent framework can be formulated in which supersymmetry is manifest even in the presence of monopoles, and this term does not contribute. Thus, the \nn supersymmetry currents given below, derived by setting this term to zero, are correct in all cases.

Let $\delta {\cal L}\equiv\partial_\mu K^\mu$. According to Noether's theorem \cite{Sri}, the first of the two supercurrents is given by 
\br
S^{(1)\mu}&=&\left[\delta^{(1)} a\,\frac{\partial{\cal L}}{\partial(\partial_\mu a)}+\delta^{(1)}\psi\,\frac{\partial{\cal L}}{\partial(\partial_\mu\psi)}+\delta^{(1)}\lambda\,\frac{\partial{\cal L}}{\partial(\partial_\mu\lambda)}\right]+c.c.\nonumber\\
&&\qquad+\delta^{(1)} A_\nu\,\frac{\partial{\cal L}}{\partial(\partial_\mu A_\nu)}-K^{(1)\mu}\nonumber\\
&=&\frac{1}{4\pi}\left\{\sqrt{2}\,({\rm Im}\,\tau)\,(\psi\,\sigma^\mu\bar\sigma^\nu\xi^{(1)})\,\partial_\nu\bar a+\frac{1}{\sqrt{2}}\,\bar\tau'(\bar\psi\bar\lambda)\,(\xi^{(1)}\sigma^\mu\bar\lambda)\right.\nonumber\\
&&-i\left.(\xi^{(1)}\sigma_\nu\bar\lambda)\,\left[({\rm Im}\,\tau)(F^{\mu\nu}-i\widetilde F^{\mu\nu})+{\rm Im}(\sqrt{2}\,\tau'\lambda\,\sigma^{\mu\nu}\psi)\right]\right\}+c.c.\label{S1}
\er
Notice that the auxiliary fields do not appear in the supercurrent. We construct the second \none  supercurrent by using the $SU(2)_R$ symmetry of the \nn  vector multiplet. In the expression (\ref{S1}) we make the transformations
$$
\xi^{(1)}\to\xi^{(2)},\qquad \lambda\to\psi,\qquad \psi\to-\lambda,
$$
($a$ and $A^\mu$ are singlets) and obtain
\brs
S^{(2)\mu}&=&\frac{1}{4\pi}\left\{-\sqrt{2}\,({\rm Im}\,\tau)\,(\lambda\,\sigma^\mu\bar\sigma^\nu\xi^{(2)})\,\partial_\nu\bar a-\frac{1}{\sqrt{2}}\,\bar\tau'(\bar\psi\bar\lambda)\,(\xi^{(2)}\sigma^\mu\bar\psi)\right.\\
&&-i\left.(\xi^{(2)}\sigma_\nu\bar\psi)\,\left[({\rm Im}\,\tau)(F^{\mu\nu}-i\widetilde F^{\mu\nu})+{\rm Im}(\sqrt{2}\,\tau'\lambda\,\sigma^{\mu\nu}\psi)\right]\right\}+c.c.
\ers
The supercurrents are each of the form
$$
S^{(a)\mu}=\xi^{(a)\alpha}S^{(a)\mu}_\alpha+\bar\xi_{\dot\alpha}^{(a)}\bar S^{(a)\mu\dot\alpha},\qquad a=1,2.
$$

In order to derive the superalgebra, we will need to express the supercurrents in terms of canonical variables. To this end, we 
compute   the following canonical momenta from the Lagrangian (2):
$$
\Pi_{a}=\frac{\p{\cal L}}{\p(\partial_0 a)}=\frac{1}{4\pi}\,({\rm Im}\,\tau)\,\partial^0\bar a,
$$
$$
(\Pi_{\psi})^\beta=(\frac{\p{\cal L}}{\p(\partial_0\psi)})^\beta=\frac{1}{8\pi}\,\bar\tau\,\bar\psi_{\dot\beta}\,(\bar\sigma^0)^{\dot\beta\beta},\quad(\Pi_{\bar\psi})_{\dot\beta}=(\frac{\p{\cal L}}{\p(\partial_0\bar\psi)})_{\dot\beta}=-\,\frac{1}{8\pi}\,\tau\,\psi^{\beta}\,(\sigma^0)_{\beta\dot\beta}
$$
$$
(\Pi_{\lambda})^\beta=(\frac{\p{\cal L}}{\p(\partial_0\lambda)})^\beta=\frac{1}{8\pi}\,\bar\tau\,\bar\lambda_{\dot\beta}\,(\bar\sigma^0)^{\dot\beta\beta},\quad(\Pi_{\bar\lambda})_{\dot\beta}=(\frac{\p{\cal L}}{\p(\partial_0\bar\lambda)})_{\dot\beta}=-\,\frac{1}{8\pi}\,\tau\,\lambda^{\beta}\,(\sigma^0)_{\beta\dot\beta}
$$
\beq\label{canmom}
\Pi_{A_\mu}=\frac{\p{\cal L}}{\p(\partial_0 A_\mu)}=-\frac{1}{4\pi}{\rm Im}\,\left[\tau\,(F^{0\mu}-i\widetilde F^{0\mu})+\sqrt{2}\,\tau'\,(\lambda\,\sigma^{0\mu}\psi)\right]
\eeq
(Notice that
$\Pi_{A_\mu}$ actually has an {\em upper} Lorentz index $\mu$.)

The two generators of \nn  supersymmetry transformations are 
\br
Q^{(1)}_\alpha&=&\int\,d\,^3x\,S^{(1)0}_\alpha({\bf x},t)\nonumber\\
&=&\int\,d\,^3x\,\left[\sqrt{2}\,\Pi_a\,\psi_\alpha+\frac{\sqrt{2}}{4\pi}\,({\rm Im}\,\tau)\,(\sigma^i\bar\sigma^0)_\alpha\,^\delta\psi_\delta\,\partial_i\bar a\right.\nonumber\\
&&-\left.\sqrt{2}\,\frac{\bar\tau'}{\bar\tau}(\bar\psi\bar\lambda)\,(\Pi_\lambda)_{\alpha}+i\left(\Pi_{A_i}-\frac{\bar\tau}{4\pi}\widetilde F^{0i}\right)(\sigma_i)_{\alpha\dot\alpha}\bar\lambda^{\dot\alpha}\right]\label{Q1}
\er
\br
Q^{(2)}_\beta&=&\int\,d\,^3y\,S^{(2)0}_\beta ({\bf x},t)\nonumber\\
&=&\int\,d\,^3y\,\left[-\sqrt{2}\,\Pi_a\,\lambda_\beta-\frac{\sqrt{2}}{4\pi}\,({\rm Im}\,\tau)\,(\sigma^j\bar\sigma^0)_\beta\,^\gamma\lambda_\gamma\,\partial_j\bar a\right.\nonumber\\
&&+\left.\sqrt{2}\,\frac{\bar\tau'}{\bar\tau}(\bar\psi\bar\lambda)\,(\Pi_\psi)_{\beta}+i\left(\Pi_{A_i}-\frac{\bar\tau}{4\pi}\widetilde F^{0i}\right)(\sigma_j)_{\beta\dot\beta}\bar\psi^{\dot\beta}\right]\label{Q2}
\er
In Appendix D, we show that these expressions reproduce the supersymmetry variation of the fields of the \nn  multiplet.

\bigskip
\section{Central Charge}

\bigskip
Using the two generators of \nn  supersymmetry transformations (\ref{Q1}-\ref{Q2}), we now calculate the central charge $Z$ from $\{Q_\alpha^{(1)},Q_\beta^{(2)}\}$. As shown in detail in Appendix E, the Dirac bracket \cite{HandT} of the two supersymmetry generators is
\beq
\left\{Q^{(1)}_\alpha,Q^{(2)}_\beta\right\}=-i(2\sqrt{2})\int\,d^3x\,(\partial_i\bar a)\left(\Pi_{A_i}-\frac{\bar\tau}{4\pi}\widetilde F^{0i}\right)\varepsilon_{\alpha\beta},\label{comm}
\eeq
where the canonical momentum $\Pi_{A_i}$ is given by (\ref{canmom}). 
The computation is quite involved, partly
because the Dirac bracket is rather complex due to the large number
of second--class constraints, as discussed in Appendix C. 

There are several conventions regarding the central charge used in the literature, which make $Z$ appear in the \nn  supersymmetry algebra preceded by various numerical factors. Our choice of conventions will ensure that 
we end up with the standard expression for $Z$
$$
Z\,=\,a_\infty\,n_e\,+\,a_{D\infty}\,n_m,
$$
where $a_\infty$ and $a_{D\infty}$ are the constant non-zero values of the field $a(x)$ and its dual field $a_D(x)$, respectively, on the sphere $r\to\infty$. The electric and magnetic charge integers are defined by
\beq
n_e=\int\,d^3x\,\partial_i\Pi_{A_i},\qquad n_m=\frac{1}{4\pi}\,\int\,d^3x\,\partial_i\,\widetilde F^{i0},\label{integers}
\eeq
Therefore, we write the realization of the \nn  algebra on massive states in the rest frame as
\begin{eqnarray}
\left\{Q_\alpha^{(a)},\bar Q_{(b)\dot\beta}\right\}&=&2\,M\,(\sigma^0)_{\alpha\dot\beta}\,\delta^a_b\label{alg1}\\
\left\{Q_\alpha^{(a)},Q_\beta^{(b)}\right\}&=&-\,i(2\sqrt{2})\varepsilon_{\alpha\beta}\,\varepsilon^{ab}\,\bar Z\label{alg2}\\
\left\{\bar Q_{\dot\alpha}^{(a)},\bar Q^{(b)}_{\dot\beta}\right\}&=&i(2\sqrt{2})\varepsilon_{\dot\alpha\dot\beta}\,\varepsilon^{ab}\,Z\label{alg3}
\end{eqnarray}
where $a, b=1,2$.

Comparing (\ref{comm}) and (\ref{alg2}) we obtain the following central charge formula
\beq\label{cc}
Z=\int\,d^3x\,(\partial_i a)\left(\Pi_{A_i}-\frac{\tau}{4\pi}\,\widetilde F^{0i}\right)
\eeq
Using $\tau=\p a_D/\p a$, we can write (\ref{cc}) as
$$
Z=\int\,d^3x\,\left[(\partial_i a)\,\Pi_{A_i}-\frac{1}{4\pi}\,(\partial_i a_D)\widetilde F^{0i}\right]
$$
Integrating by parts, we obtain
\br
Z&=&\int\,d^3x\,\left[\partial_i\left(a\,\Pi_{A_i}\right)
-\frac{1}{4\pi}\,\partial_i\left(a_D\,\widetilde F^{0i}\right)\right]
%\nonumber\\
-\int\,d^3x\,\left[a\,\partial_i\Pi_{A_i}
-\frac{1}{4\pi}\,a_D\,\partial_i\widetilde F^{0i}\right]
\nonumber\\
%&=a_\infty\oint\,ds_i\,\Pi_{A_i}-\frac{1}{4\pi}\,a_{D\infty}\oint\,ds_i\,\widetilde F^{0i}
%\nonumber\\
%&&-\int\,d^3x\,\left(a\,\partial_i\Pi_{A_i}\right)+\frac{1}{4\pi}\,\int\,d^3x\,\left(a_D\,\partial_i\widetilde F^{0i}\right)
%\nonumber\\
&&\nonumber\\
&&=a_\infty\,n_e\,+\,a_{D\infty}\,n_m\label{line1}\\
&&~~~~~~~-\int\,d^3x\,\left(a\,\partial_i\Pi_{A_i}\right)
+\frac{1}{4\pi}\,\int\,d^3x\,\left(a_D\,\partial_i\widetilde F^{0i}\right)
\label{line2}
\er
where we have replaced $a$ and $a_D$ by their asymptotic values in the
surface terms. The total central charge thus appears to differ from
the conventional value by the two integral expressions in the last
line
\footnote{When the calculation of the central charge in \cite{Iorio2} appeared, our calculation as given in 
eqs.(\ref{line1}-\ref{line2}) was substantially complete, minus our
present understanding of the terms in (\ref{line2}).}
In terms of the electric
and magnetic {\em number} densities $\nu_e$ and $\nu_m$ ({\it not} the 
charge densities), the extra integrals
can be rewritten
\beq\label{line2again}
-\int d^3x \, \left[ a({\bf x})\nu_e+a_D({\bf x})\nu_m\right]
\eeq
This integral will vanish if appropriate linear combinations of 
$a({\bf x})$ and $a_D({\bf x})$ vanish at the locations of charge sources. 
As we shall see in section 5, the BPS equations themselves imply the stronger result that $a(r)n_e+a_D(r)n_m$
%if all we have is a point dyonic charge $(n_e,n_m)$
%at the origin, then in order for the central charge to depend only 
%on the asymptotic values of the fields, we must have 
%$$
%a(0)n_e+a_D(0)n_m=0.
%$$
%As we shall see in sections 4 and 5, this condition will impose a
%significant constraint on the structure of BPS solutions. 
%In fact,
%we shall find that the BPS equations themselves imply the stronger result
% that $a\nu_e+a_D\nu_m$ 
must vanish wherever 
there is a source of electric or magnetic charge. 

As a consistency check for our identifications of electric and magnetic number densities, we conclude this section by showing how the Witten effect \cite{WE} follows from (\ref{canmom}) and (\ref{integers}). Using (\ref{canmom}) in the first of the definitions (\ref{integers}), we obtain
$$
n_e=\int\,d^3x\,\partial_i\,\Pi_{A_i}=-\frac{1}{4\pi}\,\int\,d^3x\,\partial_i\,{\rm Im}\,\left[\tau(F^{0i}-i\widetilde F^{0i})\,+\,\sqrt{2}\,\tau'(\lambda\,\sigma^{0i}\,\psi)\right]
$$
The fermionic boundary term vanishes because the fermion fields vary with distance as $\left|{\bf x}\right|^{-3/2}$). We are left with
$$
n_e=-\frac{1}{4\pi}\,\int\,d^3x\,\partial_i\,\left[({\rm Im}\,\tau)F^{0i}\right]+\frac{1}{4\pi}\,\int\,d^3x\,\partial_i\,\left[({\rm Re}\,\tau)\widetilde F^{0i}\right]
$$
On the sphere at $r\to\infty$, $\tau$ has the constant value $\tau_\infty=\tau(a_\infty)$, so we have
\begin{eqnarray*}
n_e&=&-\frac{1}{4\pi}\,({\rm Im}\,\tau_\infty)\,\int\,d^3x\,\partial_i\,F^{0i}+\frac{1}{4\pi}\,({\rm Re}\,\tau_\infty)\,\int\,d^3x\,\partial_i\,\widetilde F^{0i}\\
&=&\frac{1}{g_\infty}\,\int\,d^3x\,\partial_i\left(-\frac{F^{0i}}{g}\right)\;-\;({\rm Re}\,\tau_\infty)\,n_m\;=\;\frac{1}{g_\infty}\,\int\,d^3x\,(\partial_i\,E^i)\;-\;\frac{\theta_\infty}{2\pi}\,n_m
\end{eqnarray*}
where $\tau=\theta/2\pi\,+\,i(4\pi/g^2)$, and we have identified the electric field as $E^i=F^{i0}/g$. With the usual definition of the total electric charge,
$$
Q_e=\int\,d^3x\,\partial_i\,E^i
$$
we obtain the familiar expression of the Witten effect,
\beq\label{witten}
Q_e=\left(n_e+\frac{\theta_\infty}{2\pi}\,n_m\right)g_\infty
\eeq

\bigskip
\section {BPS Equations}

We now derive the BPS equations for the low-energy effective U(1) theory. The usual method is to write the Hamiltonian as a sum of squares and boundary terms, and to saturate the topological mass bound by setting each of the squared quantities equal to zero.  The problem with this method is that it does not tell us anything about the supersymmetry properties of solutions to the BPS equations, and in particular, it does not guarantee that these solutions will even be supersymmetric.
%This is the method originally used by Bogomol'nyi, Prasad, and Sommerfield, and it remains the only one that can be used in a non-supersymmetric context.

In this section, we will derive the BPS equations directly from supersymmetry, by requiring that the variations of the fields vanish for some linear combination of the two supersymmetry generators, as implied by the \nn  algebra. One advantage of this method is that it will tell us precisely how \nn  is broken to \none  by a given BPS state. Another advantage is that it guarantees that there are no quantum corrections to the BPS equations. 
%[too technical here] while a detailed analysis is required with the Hamiltonian method to show that terms involving derivatives of $\tau$ do not appear in the BPS equations. 
%We also hope that this method, by keeping supersymmetry in the foreground, will also eventually be useful in determining which dyon charges appear in the BPS spectrum.
\bigskip

Consider the following linear combinations of supersymmetry generators,
\begin{eqnarray}
a_{\alpha}&=&\frac{1}{\sqrt{2}}\,[Q_{\alpha}^{(1)}\,+\,\eta\,\varepsilon_{\alpha\beta}\,(\bar\sigma^0)^{\dot\gamma\beta}\,\bar Q^{(2)}_{\dot\gamma}],\label{aalpha}\\
b_{\alpha}&=&\frac{1}{\sqrt{2}}\,[Q_\alpha^{(1)}\,-\,\eta\,\varepsilon_{\alpha\beta}\,(\bar\sigma^0)^{\dot\gamma\beta}\,\bar Q^{(2)}_{\dot\gamma}].\label{balpha}
\end{eqnarray}
where $\eta$ is a complex constant whose physical significance will be explained shortly. We have
\beq
\left.
\begin{array}{lllll}
\{a_{\alpha},\bar a_{\dot\beta}\}\\
\\
\{b_{\alpha},\bar b_{\dot\beta}\}
\end{array}
\right\}
%=\frac{1}{2}\,\left( 
%\{Q_\alpha^{(1)},\bar Q^{(1)}_{\dot\beta}\}\,\pm\,\eta\,\varepsilon_{\alpha\beta}\,(\bar\sigma^0)^{\dot\gamma\beta}\{\bar Q_{\dot\gamma}^{(2)}, \bar Q^{(1)}_{\dot\beta}\}\right.\nonumber\\
%&\mp\,\bar\eta\,\varepsilon^{\gamma\delta}\{Q_\alpha^{(1)},Q^{(2)}_{\delta}\}\,(\sigma^0)_{\gamma\dot\beta}
%\\
%-\left.\left|\eta\right|^2\,\varepsilon_{\alpha\beta}\,\varepsilon^{\gamma\delta}\,(\bar\sigma^0)^{\dot\gamma\beta}\,(\sigma^0)_{\gamma\dot\beta}\,\{\bar Q_{\dot\gamma}^{(2)},Q^{(2)}_\delta\}\right)\nonumber\\
=(\sigma^0)_{\alpha\dot\beta}\,[M\,(1\,+\,\left|\eta\right|^2)\,\mp\,2\sqrt{2}\,{\rm Im}\,(\eta Z)]\label{aabb}
\eeq
where we have used eqs. (\ref{alg1}--\ref{alg3}).

Since $\{a_{\alpha},\bar a_{\dot\beta}\}$ and $\{b_{\alpha},\bar b_{\dot\beta}\}$ are semipositive definite operators, we obtain the lower mass bound
\beq\label{M}
M\geq\frac{2\sqrt{2}\,\left|{\rm Im}\,(\eta Z)\right|}{1\,+\,\left|\eta\right|^2}
\eeq
This inequality holds for all $\eta$, so it can only be
saturated for a particular $\eta$ if the right-hand side is maximized. 
This occurs precisely when 
$\eta=\pm i\bar Z/|Z|$. 
Later we will reach this same conclusion 
and determine the sign directly 
from the BPS equations of motion. 
In this case, 
eq.(\ref{M}) assumes the familiar form $M\ge \sqrt 2 |Z|$.

We wish to study to the supersymmetry properties of 
BPS states, that is, states that saturate the mass bound (\ref{M}). 
If this is true, then 
$$
\{a_{\alpha},\bar a_{\dot\beta}\}\,=\,0
$$
(taking the upper sign in (\ref{aabb}), with no loss of generality). 
%(It is irrelevant whether we choose $\{a_{\alpha},\bar a_{\dot\beta}\}$ or $\{b_{\alpha},\bar b_{\dot\beta}\}$ to vanish since this choice amounts to a redefinition $\eta\to -\,\eta$.)
In the representation of supersymmetry generators on asymptotic states, the above relation implies (by a standard argument) that $a_\alpha$ and $\bar a_{\dot\alpha}$ are projection operators onto the null state, while in the representation in which they act on fields, this property translates into
\begin{equation}\label{aphi}
\{a_\alpha,\phi\}\,=\,0,\qquad \{\bar a_{\dot\alpha}, \phi\}\,=\,0.
\end{equation}
Conversely, these relations imply that $\{a_{\alpha},\bar a_{\dot\alpha}\}\,=\,0$, {\it via} the Jacobi identity for $[a_\alpha,\{\bar a_{\dot\alpha},\phi\}]$.

Thus, BPS states are left invariant under the action of half of the supersymmetry generators, either by (\ref{aalpha}) or by (\ref{balpha}). 
For a BPS soliton made out of the component fields 
of the vector multiplet, the variations of the field configurations should 
vanish for a particular linear combination of $\xi^{(1)}Q^{(1)}$ and $\xi^{(2)}Q^{(2)}$ and their conjugates, that is, with a particular linear relation between the supersymmetry 
parameters $\xi^{(1)}$ and $\xi^{(2)}$. 
%It is important to emphasize that  choose a linear combination of generators, such as (\ref{aalpha}) and (\ref{balpha}), to obtain $\delta\phi\,=\,0$ for a BPS state. As (\ref{eq}) clearly shows, a condition regarding the parameters of the supersymmetry transformation must also be imposed. 

To find this linear relation, let us first re-express a general \nn 
supersymmetry variation of a component field $\phi$ in terms of 
the generators $a$ and $b$: 
\br
\delta\phi&=&i\xi^{(1)\alpha}\{Q_\alpha^{(1)},\phi\}\, +\,i\bar\xi^{(1)}_{\dot\alpha}\{\bar Q^{(1)\dot\alpha},\phi\}\, +\,i\xi^{(2)\beta}\{Q_\beta^{(2)}, \phi\}\,+\,i\bar\xi^{(2)}_{\dot\beta}\{\bar Q^{(2)\dot\beta}, \phi\}\nonumber\\
&= &\frac{i}{\sqrt{2}}\,[(\xi^{(1)\alpha}\, -\,\eta^{-1}\,\bar \xi^{(2)}_{\dot\beta}\,(\bar\sigma^0)^{\dot\beta\alpha})\,\{a_\alpha,\phi\}
+\,(-\,\bar\xi^{(1)\dot\alpha}\,+\,\bar\eta^{-1}\,(\bar\sigma^0)^{\dot\alpha\alpha}\,\xi^{(2)}_\alpha)\,\{\bar a_{\dot\alpha}, \phi\}\nonumber\\
&+&\!\!\!(\xi^{(1)\alpha}\,+\,\eta^{-1}\,\bar\xi^{(2)}_{\dot\beta}\,(\bar\sigma^0)^{\dot\beta\alpha})\,\{b_\alpha,\phi\}
%\nonumber\\
+\,(-\,\bar\xi^{(1)\dot\alpha}\, -\,\bar\eta^{-1}\,(\bar\sigma^0)^{\dot\alpha\alpha}\,\xi^{(2)}_\alpha)\,\{\bar b_{\dot\alpha}, \phi\}] \label{eq}
\er
where $\phi\,=\,a,\,\lambda_\alpha,\,\psi_\alpha,\,A_\mu$ and the curly brackets denote commutators or anticommutators as appropriate.
Now using (\ref{aphi}) in (\ref{eq}) we obtain
\begin{eqnarray*}
\delta\phi&=&\frac{i}{\sqrt{2}}\,\left[\left(\xi^{(1)\alpha}\,+\,\eta^{-1}\,\bar\xi^{(2)}_{\dot\beta}\,(\bar\sigma^0)^{\dot\beta\alpha}\right)\,\{b_\alpha,\phi\}+\left(-\,\bar\xi^{(1)\dot\alpha}\, -\,\bar\eta^{-1}\,(\bar\sigma^0)^{\dot\alpha\alpha}\,\xi^{(2)}_\alpha\right)\,\{\bar b_{\dot\alpha},\phi\}\right]
\end{eqnarray*}
This identifies the particular supersymmetry transformation which leaves the fields of a BPS solution invariant, $\delta\phi\,=\,0$; it is the transformation for which the parameters satisfy the relation
\begin{equation}\label{xi}
\xi^{(1)\alpha}\,=\,-\,\eta^{-1}\,\bar\xi^{(2)}_{\dot\beta}\,(\bar\sigma^0)^{\dot\beta\alpha}
\end{equation}

We will now derive the BPS equations for a bosonic field configuration by using (\ref{xi}) in the explicit expression for $\delta\phi\,=\,0$, thereby saturating the mass bound (\ref{M}).

The \nn  variation of $\lambda_\alpha$ is
$$
\delta\lambda_\alpha\,=\,-\,(\sigma^{\mu\nu})_\alpha\,^\beta\,\xi_\beta^{(1)}\,F_{\mu\nu}\,-\,i\sqrt{2}\,(\sigma^\mu)_{\alpha\dot \beta}\,\bar \xi^{(2)\dot\beta}\,\partial_\mu a
$$
Notice that the auxiliary fields $f$ and $D$ do not appear in this expression because, as seen in section 2, their constraint equations involve fermion fields, which we have set equal to zero. Since $\sigma^{\mu\nu}F_ {\mu\nu}=-(\sigma^i\bar\sigma^0)(F_{0i}-i\widetilde F_{0i})$, we have
\brs
\delta\lambda_\alpha&=&(\sigma^i\bar\sigma^0)_\alpha\,^\beta\,\xi_\beta^{(1)}\,(F_{0i}-i\widetilde F_{0i})-i\sqrt{2}\,(\sigma^0)_{\alpha\dot \beta}\,\bar \xi^{(2)\dot\beta}\,\partial_0 a-i\sqrt{2}\,(\sigma^i)_{\alpha\dot \beta}\,\bar\xi^{(2)\dot\beta}\,\partial_i a\\
&=&\left[(\sigma^i)_{\alpha\dot\beta}\,\left((F_{0i}-i\widetilde F_{0i})-i\sqrt{2}\eta\,\partial_i a\right)+(\sigma^0)_{\alpha\dot\beta}\,(\,-\,i\sqrt{2}\eta\,\partial_0 a)\right]\,(\bar\sigma^0)^{\dot\beta\beta}\,\xi_\beta^{(1)}
\ers
using (\ref{xi}). 
%Here we see once again that the existence of the relation (\ref{xi}) between supersymmetry parameters is essential in the derivation of the BPS equations. 
%Without such a relation, it would have been impossible to form a common factor in the above expression and bring $F_{\mu\nu}$ and $\partial_\mu a$ inside the same parenthesis. 
In order for $\delta\lambda_\alpha\,=\,0$ for any transformation parameter $\xi_\beta^{(1)}$, the square parenthesis in the expression above must vanish. 
Since the matrices $(\sigma^0, {\bf\sigma})$ are linearly independent, the following equations must hold
\begin{eqnarray}
F^{0i}-i\widetilde F^{0i}&=&i\sqrt2\eta\,\partial^ia,\label{BPS}\\
\partial_0a&=&0.\label{bps}
\end{eqnarray}
This is the second pair of BPS equations of the U(1) effective theory. Eq. (21) shows that for any BPS state the scalar field $a$ is static. 

Applying now the same treatment to the \nn   variation of $\psi_\alpha$, we obtain
\begin{eqnarray}
F^{0i} - i\widetilde F^{0i}&=&i\sqrt2\bar\eta^{-1}\,\partial^ia,\label{BPS2}\\
\partial_0a&= &0.
\end{eqnarray}
Eq. (\ref{BPS2}) agrees with (\ref{BPS}) if $\bar\eta=\eta^{-1}$, i.e. if $\eta$ is a phase factor, $\left|\eta\,\right|^2=1$.
The mass bound (\ref{M}) becomes
\begin{equation}\label{M2}
M\geq\sqrt{2}\left|{\rm Im}\,(\eta Z)\right|
\end{equation}

We have explained previously that $\eta=\pm i\bar Z/|Z|$. 
To determine the sign of $\eta$, we now
use the definition (\ref{canmom}) of the 
canonical momentum $\Pi_{A_i}$ to rewrite Eq. (\ref{BPS}) as
\begin{equation}\label{BPS3}
\Pi_{A_i}\,-\,\frac{\bar \tau}{4\pi}\,\widetilde F^{0i}\,=\,-\,\frac{i\sqrt2\,\eta}{4\pi}\,({\rm Im}\,\tau)\,\partial^ia
\end{equation} 
%We can obtain a useful integral expression for the central charge $Z$ in terms of only the scalar field $a$. Indeed,
By taking the complex conjugate of Eq. (\ref{BPS3}), contracting it with $\partial_ia$, integrating over all space and comparing with (\ref{cc}), we get a nice expression for the central charge $Z$:
\br
Z&= &\int\,{d\,}^3x\,(\partial_ia)(\Pi_{A_i}\,-\,\frac{\tau}{4\pi}\,\widetilde
F^{0i})\nonumber\\
&=&\frac{i\sqrt2\bar\eta}{4\pi}\,\int\,{d\,}^3x\,({\rm Im}\,\tau)\,\partial_ia\,\partial^i\bar a\label{Z2}
\er
%From Eq. (\ref{Z2}) we can immediately read off the significance of
%$\eta$. We have already seen that $\eta$ is just a phase factor. Then,
%Eq. (\ref{Z2}) shows that $-i\bar\eta$ is the phase of the central
%charge $Z$, or
This shows that $-i\bar\eta$ is the phase of the central charge $Z$,
\beq\label{eta}
\eta=-i\,{\bar Z\over\left|Z\right|},
\eeq
so the mass bound (\ref{M2}) becomes
\begin{equation}\label{M3}
M\geq\sqrt{2}\,\left|Z\right|.
\end{equation} 

We can obtain an equation involving only $a$ by taking the 
imaginary part of the 
divergence of the BPS equations in the form of eq.(\ref{BPS}): 
\beq\label{harmsource}
\nabla^2 \re (\eta\, a) = -{4\pi\over \sqrt2} \,\nu_m
\eeq

Multiplying eq.(\ref{BPS3}) through by $\tau$ and taking the divergence of its imaginary part gives
\beq\label{harmsourceD}
\nabla^2 \re (\eta\, a_D) = {4\pi\over \sqrt2} \,\nu_e
\eeq
We thus have two harmonic equations for two independent real 
scalars $X\equiv\re (\eta\, a)$ and $X_D\equiv\re(\eta\, a_D)$. 

As usual, satisfaction of the BPS equations automatically
implies that the equations of motion are satisfied. 
To see this, 
we rewrite the equation of motion for $a$
eq.(\ref{eqfora}) in the form
$$
\partial_{\mu}\tau\partial^{\mu}a+2i({\rm Im}\,\tau)\partial_\mu\partial^\mu a-\frac{1}{2}\,\bar\tau'(F_{0i}+i\widetilde F_{0i})(F^{0i}+i\widetilde F^{0i})=0,
$$
for a bosonic BPS state. The BPS equations may be used to 
eliminate the gauge field strength and obtain an equation involving only $a$:
$$
\tau'\partial_i a\partial^i a-2i({\rm Im}\,\tau)\nabla^2a+\bar\eta^2\bar\tau'\partial_i \bar a\partial^i \bar a=0
$$
Multiplying the above equation by $\eta$ and taking the
imaginary part gives
%$$
%\eta\,\tau'\partial_i a\partial^i a-2i\eta\,({\rm Im}\,\tau)\nabla^2a+\bar\eta\,\bar\tau'\partial_i \bar a\partial^i \bar a=0
%$$
%whose imaginary part gives 
the source-free version of eq.(\ref{harmsource}),
\begin{equation}\label{harma}
\nabla^2{\rm Re}\left(\eta a\right)=0
\end{equation}
after dividing by ${\rm Im}\,\tau$. 
%(A magnetic source as in 
%eq.(\ref{xsoln}) at a point where ${\rm Im}\,\tau=0$ is
%allowed by this equation.)
The real part gives 
\beq\label{extracond}
\re (\eta\,\partial_i \tau\partial^i a)+(\im\tau)\nabla^2(\im\eta a)=0
\eeq
which after a couple of lines of algebra becomes
%$$
%\re (\eta\,\partial_i \tau\partial^i a)+(\re\tau)(\re\eta\nabla^2a)-\re(\tau\eta\nabla^2a)=0.
%$$
%But
%$$
%\re(\tau\eta\nabla^2a)=\re(\eta\nabla^2a_D)+\re (\eta\,\partial_i \tau\partial^i a)
%$$
%so (\ref{extracond}) becomes
$$
\nabla^2\re(\eta a_D)-(\re\tau)\nabla^2(\re\eta a)=\re(\eta\nabla^2a_D)=0,
$$
using (\ref{harma}).
%which is trivially satisfied by (\ref{harma}) and (\ref{harmaD}).

The equation for motion for $A^\mu$, 
eq.(\ref{eqgauge}), contains no further information.
After setting fermion fields to zero, it reduces to
$$
\partial_i{\rm Im}\left[\tau(F^{0i}-i\widetilde F^{0i})\right]=0
$$
Substituting now the BPS equation (\ref{BPS}), we  obtain the source-free version of eq.(\ref{harmsourceD}).
\beq\label{harmaD}
\nabla^2 {\rm Re}\left(\eta a_D\right)=0
\eeq
In order to include a source of electric or magnetic charge
on the right--hand sides of (\ref{harma}-\ref{harmaD}), as in eqs.(\ref{harmsource}-\ref{harmsourceD}),
we may add a source term to the Lagrangian as
will be discussed in Section 6. 

A solution for 
a localized spherically symmetric source of total magnetic charge $n_m=1$ and $n_e=0$
may   readily be found.
Away from any sources, the solution looks like 
\br
X(r)&=&X_\infty + {1\over \sqrt 2\, r}\label{Xsolution}\\
X_D(r)&=& 0\label{XDsolution}
\er
where $X_\infty \equiv \re (\eta\, a_\infty)$ is the asymptotic value of $X$. 
The second equation in particular implies that $\eta\, a_D(r)$ is purely imaginary for all $r$, or in other words that the phase of $a_D$ is constant: 
\beq\label{constphase}
{a_D(r)\over |\,a_D(r)|}= \pm i\bar\eta  \qquad\qquad \hbox{for all $r$.}
\eeq
At $r=\infty$, we have determined that the right-hand side is $-i\bar\eta$. 
If $a_D(r)$ passes through zero at some finite $r=r_0$, the sign will flip. 
The constant phase of $a_D(r)$ (up to a sign) 
and the semi-infinite range of $a$ 
makes it plausible that this occurs, for a wide range of 
boundary values $a_\infty$. 
In fact, we will argue in the next section that such an $r_0$ exists, 
for {\it any} value of $a_\infty$.

More generally, the solution for a dyon of charges $(n_m,n_e)$ is
\br\label{xsoln}
X(r)&=&X_\infty + {n_m\over \sqrt 2\, r}\label{Xdyonsolution}\\
X_D(r)&=& X_{D\infty} - {n_e\over \sqrt 2\, r}\label{XDyonsolution}
\label{xsolnd}\er
Note that the ``local central charge'' \cite{CRV}
$$
Z(r)=n_e\,a(r)\,+\,n_m\,a_D(r)
$$
has constant phase $-i\bar \eta$. Just as in the case of a simple monopole, we conclude that there exists a radius $r_0$, defined by the duality-invariant quantity $Z_0=Z(r_0)=a_0n_e+a_{D0}n_m=0$, for which the local central charge vanishes. This statement will play an important role in subsequent sections. 

It is also possible to solve 
eqs.(\ref{harmsource}--\ref{harmsourceD}) 
directly for $a(r)$. 
Take as an ansatz a spherically symmetric
harmonic function 
\beq\label{aansatz}
a(r)=a_\infty+{C\over \sqrt2 r}
\eeq
This satisfies eq.(\ref{harmsource}) with a delta-function
source at the origin provided that
\beq\label{etac}
\re (\eta\, C)=n_m
\eeq
The left side of 
eq.(\ref{harmsourceD}) can be rewritten in terms of $a$ as
\br
\nabla^2 \re (\eta\, a_D) &=& \re (\eta\,\tau\nabla^2 a)-\re\left( \eta\,(\partial_i \tau)(\partial^i a)\right)\\
&=&\re(\tau) \re(\eta\,\nabla^2a)-
\im(\tau)\im(\eta\,\nabla^2a)
-\re\left(\eta\,\partial_i\left({da_D\over da}\right)\partial^i a\right)
\nonumber
\er
The first two terms on the right vanish if $\tau$ vanishes at the location of the source, as will be shown in Section 5. The third term vanishes by eq.(\ref{constphase}) and because $\partial^i a$ has constant phase $C$, and we recover eq.(\ref{harmaD}). 

The integration constant $C$ can be fixed by demanding, in addition to eq.(\ref{etac}), that
our solution look like a magnetic monopole at infinity. 
At large $r$, we require
\br
{\bf E}&\sim&\frac{Q_e}{4\pi}\,\frac{\hat r}{r^2}\;=\;\left(n_e+\frac{\theta_\infty}{2\pi}\,n_m\right)\,{g_\infty\over4\pi}\,\frac{\hat r}{r^2},\label{E}\\
{\bf B}&\sim&\frac{Q_m}{4\pi}\,\frac{\hat r}{r^2}\;=\;\frac{n_mg_{m\infty}}{4\pi}\,\frac{\hat r}{r^2}\;=\;\frac{n_m}{g_\infty}\,\frac{\hat r}{r^2},\label{B}
\er
where we have used (\ref{witten}) in (\ref{E}) and $g_{m\infty}=4\pi/g_\infty$ in (\ref{B}). 
The electric and magnetic fields are defined by 
$$
E^i\,=\,\frac{F^{i0}}{g}\qquad B^i\,=\,\frac{\widetilde F^{i0}}{g},
$$
in terms of which the BPS equation (\ref{BPS}) becomes
%$$
%E^i\,-\,i\,B^i\,=\,-\,\frac{i\sqrt2\eta}{g}\,\partial^ia
%$$
%or
\beq\label{BPS4}
{\bf E}-i\,{\bf B}=\frac{i\sqrt{2}}{g}\,\eta\,{\bf\nabla} a.
\eeq

By substituting (\ref{E}-\ref{B}) and (\ref{aansatz}) into (\ref{BPS4}), we find
\beq\label{reetac}
{\rm Re}(\eta\,C)=n_m,
\eeq
and
\beq\label{imetac}
{\rm Im}(\eta\,C)=\left(n_e+\frac{\theta_\infty}{2\pi}\,n_m\right)\,{g_\infty^2\over4\pi}
\eeq
%Using (\ref{eta}), we now obtain 
We thus find
\beq\label{C}
C=i\bar\eta\,{n_e+\bar\tau_\infty n_m\over\im\tau_\infty},
\eeq
In particular, for a simple monopole $(n_e=0,\;n_m=1)$,
\beq\label{Cmono}
a(r)=a_\infty+{i\bar \eta\over \sqrt 2}\,{\bar\tau_\infty\over{\rm Im}\,\tau_\infty}\,{1\over r}.
\eeq

\bigskip
We would now like to discuss the duality transformation 
properties of the BPS equations and their solutions, and
of the full equations of motion. 
%The BPS equations (\ref{harmsource}) and
%(\ref{harmsourceD}) are manifestly duality invariant, 
%so any duality transformation of a solution is still 
%a solutions. For example, consider the effect of the
%$T$-transformations on our monopole solution
%with $\theta=0$. Since $T$ leaves $a(r)$ unchanged,
%eq.(\ref{Cmono}) is still valid. In this way we can
%obtain $a(r)$ for any dyon with unit magnetic charge. 
%It is well known these dyons comprise the complete 
%set of stable BPS solitons for the SU(2) effective 
%theory. 
Under a general duality transformation, implemented by the action of the group $PSL(2,\Z)$ (see Appendix F), we have
$$
\tau\to\frac{\alpha\tau+\beta}{\gamma\tau+\delta},\quad{\rm where~}\left(\begin{array}{cc}
\alpha&\beta\\
\gamma&\delta
\end{array}\right)\;\in\;PSL(2,\Z),
$$$$
{\bf E}-i\,{\bf B}\to e^{i\varphi}({\bf E}-i\,{\bf B}),\quad{\rm where~}e^{i\varphi}\,=\,\frac{\gamma\tau+\delta}{\left|\gamma\tau+\delta\right|},
$$$$
\partial_\mu\tau\to\frac{\partial_\mu\tau}{(\gamma\tau+\delta)^2},\quad\partial_\mu a\to(\gamma\tau+\delta)\partial_\mu a,
$$$$
{\rm Im}\,\tau\to\frac{{\rm Im}\,\tau}{\left|\gamma\tau+\delta\right|^2}.
$$
The duality transformation of eq.(\ref{BPS4}) gives
$$
e^{i\varphi}(E^i\,-\,i\,B^i)\,=\,-\,\frac{i\sqrt2\eta}{g}\,\frac{\gamma\tau+\delta}{\left|\gamma\tau+\delta\right|}\,\partial^ia
$$
so the BPS equation is duality-invariant. Eq.(\ref{Z2}) is also manifestly invariant under duality transformations. There we only need to consider the integrand, for which we have
\brs
({\rm Im}\,\tau)\,\partial_ia\,\partial^i\bar a&\to&\frac{{\rm Im}\,\tau}{\left|\gamma\tau+\delta\right|^2}\,(\gamma\tau+\delta)\partial_ia\,(\gamma\bar\tau+\delta)\partial^i\bar a\\
&&=({\rm Im}\,\tau)\,\partial_ia\,\partial^i\bar a
\ers
By contrast, the full equations of motion of the present theory are not duality-invariant in general. Indeed, let us consider the transformation properties of eq. (\ref{eqfora}), which can be written as
$$
\partial_{\mu}\tau\partial^{\mu}a+2i({\rm Im}\,\tau)\partial_\mu\partial^\mu a+\frac{1}{2}\,\bar\tau'({\bf E}+i{\bf B})\cdot({\bf E}+i{\bf B})=0
$$
The duality-transformed equation is
\brs
&&\partial_{\mu}\tau\partial^{\mu}a\left[\frac{1}{\gamma\tau+\delta}+\frac{2i({\rm Im}\,\tau)\gamma}{\left|\gamma\tau+\delta\right|^2}\right]+\frac{2i({\rm Im}\,\tau)}{\gamma\bar\tau+\delta}\pd\mu\pu\mu a\\
&&\qquad+\frac{1}{2}\,\bar\tau'\,\left(\frac{e^{-i\varphi}}{\gamma\bar\tau+\delta}\right)^2({\bf E}+i{\bf B})\cdot({\bf E}+i{\bf B})=0
\ers
Multiplying by $\gamma\bar\tau+\delta$\,($\ne0$), we obtain
$$
\partial_{\mu}\tau\partial^{\mu}a+2i({\rm Im}\,\tau)\partial_\mu\partial^\mu a+\frac{1}{2}\,\frac{1}{\gamma\tau+\delta}\,\bar\tau'({\bf E}+i{\bf B})\cdot({\bf E}+i{\bf B})=0
$$
Hence the equation of motion is invariant only if $\gamma=0,\,\delta=1$. The condition that the determinant of the transformation matrix is equal to 1 further implies that $\alpha=1$. We recover then the well known fact \cite{SW1} that the theory is not invariant under a general duality transformation (for example, it is not S-invariant), but only under a subgroup of $PSL(2,\Z)$ consisting of the matrices
$$
\left(\begin{array}{cc}
1&\beta\\
0&1
\end{array}\right)
$$
The elements of a symmetry group of an equation transform solutions of the equation into other solutions. This symmetry group preserves $n_m$,
$$
(n_m,n_e)\to(n_m,n_e-n_m\beta)
$$
Since a magnetic monopole $(1,0)$ is a solution of the equations of motion, it follows that all dyons $(1,-\beta)$ are also solutions, with the same $a$ and with $a_D\to a_D+\beta a$. Therefore, given a monopole solution, we can construct a dyon solution of unit magnetic charge by T-duality. (The fact that under this transformation $a(r)$ remains the same can also be seen directly from its expression.)
%, in which $C$ is T-duality invariant
%Not surprisingly, our radial solution $a(r)$ can in fact work only for these dyons, because the dyons with magnetic charge $\pm 1$ are the only ones with spherical symmetry (all others are multi-centered field configurations).

\bigskip
We have seen in this section that the derivation of the BPS equations from the \nn  supersymmetry algebra guarantees that they do not receive any quantum corrections, their structure being dictated solely by the supersymmetry variations of the fields contained by the \nn  vector multiplet, without reference to a specific Lagrangian. More precisely, the form of the Lagrangian can affect (\ref{bps}) through the auxiliary fields (an example of this will be seen in the next section), but cannot affect the form of (\ref{BPS}).

\bigskip

\section{The Magnetic Monopole and the Dual Lagrangian}

\bigskip
We saw in the previous section that the BPS equations derived from the
Seiberg--Witten low-energy Lagrangian lead to monopole solutions which
are singular at the origin. In fact, this Lagrangian breaks down as
$r\to 0$, for two reasons: the higher--order terms in the derivative
expansion become important at short distances $< 1/\Lambda$ and the
magnetic monopole state becomes massless at the radius $r_0$ at which
$a_D(r_0)=0$.  Near $r_0$, it is appropriate instead to work with a
dual Lagrangian containing, in addition to the dual vector multiplet, an electrically coupled hypermultiplet representing the monopole.  Our
discussion will be valid when $r_0\gg 1/\Lambda$.  It is easy to see
that the size of the monopole grows without limit as $a_{D\infty}\to 0$ (i.e., as the vevs of $a$ and $a_D$ approach their values at 
the monopole point). Indeed, from eq.(\ref{aansatz}), we obtain
that $r_0=(C/\sqrt 2)[a(r_0)-a_\infty]^{-1}$. 

In the (S-transformed) dual theory, a magnetic monopole appears as an elementary matter field, electrically charged, coupled to the dual gauge field $A^\mu_D$. (In contrast, a magnetic charge cannot couple locally to the gauge field $A^\mu$ of the original Lagrangian.) This elementary monopole field will be the scalar component of a chiral superfield. To describe such a charged massive field, we must include both its left and right chiral components and the left and right chiral components of the anti-monopole. Therefore, the Lagrangian of the theory will contain two (left) chiral superfields, ${\cal M}$ and $\widetilde{\cal M}$, which couple to the vector superfield with opposite charges. To be specific, the underlying dual Lagrangian ${\cal L}_D$ contains the S-transformed of Lagrangian (1) (with the chiral superfield $\Phi$ replaced by $\Phi_D$ and the vector superfield $W$ replaced by $W_D$), the canonical kinetic terms for ${\cal M}$ and $\widetilde{\cal M}$, and the superpotential, uniquely determined by \nn  supersymmetry as
$$
\sqrt{2}\,\Phi_D{\cal M}\widetilde{\cal M}\,+\,c.c.
$$
In terms of \none  superfields, the Lagrangian is
\brs
{\cal L}_D&=&\frac{1}{4\pi}\,Im\left[\int d^2\theta\int d^2\bar\theta\,{\p{\cal F_D}\over\p\Phi_D}\,\bar\Phi_D\,+\,{1\over 2}\,\int d^2\theta\,{\pu 2{\cal F_D}\over\p\Phi_D^2}\,W_D^\alpha W_{D\alpha}\right]\\
&&+\int d^2\theta\int d^2\bar\theta\,\left[{\cal M}^\dagger e^{2V_D}{\cal M}+\widetilde{\cal M}^\dagger e^{-2V_D}\widetilde{\cal M}\right]\\
&&+\left(\int d^2\theta\,\left[\sqrt{2}\,\widetilde{\cal M}\Phi_D{\cal M}\right]+c.c.\right)
\ers
The field content of the superfield ${\cal M}$ is given in Appendix A. With the notation given there, the bosonic part\footnote{In this section we refer only to bosonic BPS states, and ignore the fermionic part of ${\cal L}_D$.} of ${\cal L}_D$ is
\begin{eqnarray*}
{\cal L}_D^{(bosonic)}\!\!&=&\!\!\!\frac{1}{4\pi}\,{\rm Im}\,\left[\tau_D\,\left(\partial_\mu a_D \partial^\mu\bar a_D\,+\,f_D\bar f_D\,-\,\frac{1}{4}\,F_{D\mu\nu}(F^{\mu\nu}_D-i\widetilde F^{\mu\nu}_D)+{1\over2}D_D^2\right)\right]\\&&\!\!\!\!\!+\,(D_\mu M)^\dagger (D^\mu M)\,+\,(D_\mu\widetilde M)^\dagger (D^\mu\widetilde M)\,+\,f_M\bar f_M\,+\,f_{\widetilde M}\bar f_{\widetilde M}\\&&\!\!\!\!\!+(M^\dagger M-\widetilde M^\dagger \widetilde M)D_D\,+\,\sqrt{2}\,\left[\,a_D\left(f_M\widetilde M+f_{\widetilde M}M\right)\,+\,f_D\widetilde MM\right]\\&&\!\!\!\!\!+\,\sqrt{2}\,\left[\,\bar a_D\left(\bar f_M\widetilde M^\dagger +\bar f_{\widetilde M}M^\dagger \right)\,+\,\bar f_D\widetilde M^\dagger M^\dagger \right],
\end{eqnarray*}
where $D^\mu=\partial^\mu+iA^\mu_ D$. Eliminating the auxiliary fields using their constraint equations,
\begin{eqnarray*}
f_D&\!=\!&-\frac{4\pi}{{\rm Im}\,\tau_D}\,\sqrt{2}\,\widetilde M^\dagger M^\dagger ,\quad\bar f_D\;=\;-\frac{4\pi}{{\rm Im}\,\tau_D}\,\sqrt{2}\,\widetilde MM,\\
f_M&\!=\!&-\sqrt{2}\,\bar a_D\widetilde M^\dagger ,\quad\bar f_M\;=\;-\sqrt{2}\,a_D\widetilde M,\\
f_{\widetilde M}&\!=\!&-\sqrt{2}\,\bar a_DM^\dagger ,\quad\bar f_{\widetilde M}\;=\;-\sqrt{2}\,a_DM,\\
D_D&\!=\!&-\frac{4\pi}{{\rm Im}\,\tau_D}\,(M^\dagger M-\widetilde M^\dagger \widetilde M),
\end{eqnarray*}
we obtain
\begin{eqnarray*}
{\cal L}_D^{(bosonic)}&=&\frac{1}{4\pi}\,{\rm Im}\,\left[\tau_D\,\left(\partial_\mu a_D \partial^\mu\bar a_D\,-\,\frac{1}{4}\,F_{D\mu\nu}(F^{\mu\nu}_D-i\widetilde F^{\mu\nu}_D)\right)\right]\\
&&+(D_\mu M)^\dagger (D^\mu M)+(D_\mu\widetilde M)^\dagger (D^\mu\widetilde M)\,-\,2\,\left|\,a_D\right|^2(M^\dagger M\,+\,\widetilde M^\dagger \widetilde M)\\
&&-\,{1\over2}\,\frac{4\pi}{{\rm Im}\,\tau_D}(M^\dagger M\,+\,\widetilde M^\dagger \widetilde M)^2
\end{eqnarray*}
Expanding around the asymptotic values of the fields,
the second line of the above expression contains 
a mass term for the monopole field, 
with mass $\sqrt{2}\,\left|\,a_{D\infty}\right|$, 
in agreement with the BPS mass formula. 

To obtain the modifications of the BPS equations 
due to $M$, we consider as before
the \nn  supersymmetry variation of $\lambda_{D\alpha}$
\brs
\delta\lambda_{D\alpha}&=&-\,(\sigma^{\mu\nu})_\alpha\,^\beta\,\xi_\beta^{(1)}\,F_{D\mu\nu}+i\xi^{(1)}_\alpha D_D\\
&&-i\sqrt{2}\,(\sigma^\mu)_{\alpha\dot \beta}\,\bar \xi^{(2)\dot\beta}\,\partial_\mu a_D-\sqrt2\xi^{(2)}_\alpha\bar f_D
\ers
Using (\ref{xi}) we find
\brs
\delta\lambda_{D\alpha}&=&\left[(\sigma^i)_{\alpha\dot\beta}\left((F_{D0i}-i\widetilde F_{D0i})-i\sqrt{2}\eta\,\partial_i a_D\right)\right.\\
&&+\left.(\sigma^0)_{\alpha\dot\beta}\left(iD_D-i\sqrt{2}\eta\,\partial_0 a_D\right)\right](\bar\sigma^0)^{\dot\beta\beta}\xi_\beta^{(1)}-\sqrt2\xi^{(2)}_\alpha\bar f_D
\ers
With the same reasoning as for $\delta \lambda=0$ in section 4, we conclude that for $\delta\lambda_{D\alpha}=0$, the coefficients of $\xi^{(1)}$ and $\xi^{(2)}$ must vanish independently, giving
\br
F^{0i}_D-i\widetilde F^{0i}_D&=&i\sqrt2\eta\,\partial^ia_D,\label{BPSD}\\
iD_D&=&i\sqrt{2}\eta\,\partial_0 a_D,\nonumber\\
f_D&=&0.\nonumber
\er

These equations are easily solved. 
Taking the imaginary part of the divergence of the first equation
gives 
$$
\nabla^2 \,\re(\eta\, a_D) =0
$$
implying that $a_D$ has constant phase as we found earlier.
The last equation, $f_D=0$, 
implies that either $M$ or $\widetilde M$ is equal to zero, 
at least wherever $\im \tau_D\ne \infty$. 
For a static solution, we also must have 
$D_D=0$, which means, when $\im \tau_D\ne \infty$,
that $|M|=|\widetilde M|$. Now $\im \tau_D= \infty$
when $\im \tau=0$, which occurs precisely when $a_D=0$
\cite{SW1}. 
%(Both occur at the monopole point, because both $a_D=0$ 
%and $\tau=0$ are fixed points of the same 
%$SL(2,\Z)$ monodromy generator.) 
Thus, in a BPS soliton, $M$ can  only be nonzero 
where $a_D=0$. As we will discuss shortly,
this result is consistent with our 
observation in Section 3 that in order for the central
charge to agree with the usual expression
the integral of $a_D \nu_m$ must vanish. 

There are additional BPS equations involving $M$ to be 
derived by examining the \nn  transformations of the
fermionic fields in the $({\cal M, \widetilde M})$ 
hypermultiplet. The \none  variation of the \none  superpartner of 
$M$, $\psi_M$, is 
\brs
\delta^{(1)}\psi_{M\alpha}&=&i\sqrt2(\sigma^\mu)_{\alpha\dot\alpha}\bar\xi^{(1)\dot\alpha}D_\mu M+\sqrt2\xi^{(1)}_\alpha f_M\\
&=&i\sqrt2(\sigma^\mu)_{\alpha\dot\alpha}\bar\xi^{(1)\dot\alpha}D_\mu M-2\xi^{(1)}_\alpha\bar a_D\widetilde M^\dagger
\ers
When $SU(2)_R$ acts on the \nn  hypermultiplet containing ${\cal M}$ and $\widetilde{\cal M}$, it transforms $M\to\widetilde M^\dagger$, leaving $\psi_{M\alpha}$ and $\bar\psi_{\widetilde M\dot\alpha}$ invariant \cite{SW1}.
Thus, by  applying $SU(2)_R$ to the preceding transformation, 
we get
$$
\delta^{(2)}\psi_{M\alpha}=i\sqrt2(\sigma^\mu)_{\alpha\dot\alpha}\bar\xi^{(2)\dot\alpha}D_\mu\widetilde M^\dagger-2\xi^{(2)}_\alpha\bar a_D M 
$$
It follows that the full \nn  variation of $\psi_{M\alpha}$ is
\brs
\delta\psi_{M\alpha}&=&i\sqrt2(\sigma^\mu)_{\alpha\dot\alpha}\bar\xi^{(1)\dot\alpha}D_\mu M-2\xi^{(1)}_\alpha\bar a_D\widetilde M^\dagger\\
&&+i\sqrt2(\sigma^\mu)_{\alpha\dot\alpha}\bar\xi^{(2)\dot\alpha}D_\mu\widetilde M^\dagger-2\xi^{(2)}_\alpha\bar a_D M
\ers
Using again (\ref{xi}), we have
\brs
\delta\psi_{M\alpha}&=&i\sqrt2\left[(\sigma^i)_{\alpha\dot\alpha}D_i M+(\sigma^0)_{\alpha\dot\alpha}\left(D_0 M-i\sqrt2\,\bar\eta\,\bar a_D M\right)\right]\bar\xi^{(1)\dot\alpha}\\
&&+i\sqrt2\left[(\sigma^i)_{\alpha\dot\alpha}D_i\widetilde M^\dagger+(\sigma^0)_{\alpha\dot\alpha}\left(D_0\widetilde M^\dagger+i\sqrt2\,\bar\eta\,\bar a_D\widetilde M^\dagger\right)\right]\bar\xi^{(2)\dot\alpha}
\ers
Then from $\delta\psi_{M\alpha}=0$ we obtain the following additional BPS equations:
$$
D_i M=0,\quad D_i\widetilde M=0,
$$$$
D_0 M-i\sqrt{2}\,\bar\eta\,\bar a_DM=0,\quad D_0\widetilde M-i\sqrt{2}\,\eta\,a_D\widetilde M=0
$$
For an electrostatic field, we can choose 
$A_D^0$ to be the only non-zero component of the dual gauge field. The last two equations imply that for a static field configuration we have $M=0$ and $\widetilde M=0$ as long as $A_0\ne 0 $ and $a_D\ne 0$. This is certainly consistent with our earlier conclusion that $M=0$ unless $a_D=0$. 
Taking into account the first 
two equations, which imply that the fields $M$ and $\widetilde M$ are constant throughout space, we might conclude that since
$\langle M\rangle =\langle\widetilde M\rangle=0$ everywhere, the field $M$ cannot act as a source of charge. However, if we are interested in electrically charged excitations of the field $M$, we can use the $M$ operator to create any desired charge distribution, without violating the field equations of $M$. 
%we should allow for discontinuities in the $M$ field where charge is localized (just as we do in QED, where charge creation at a point produces a delta function source for

The electric charge density is given by the time component of 
the Noether current associated with U(1) gauge invariance
\beq\label{j0}
\nu_e^{\,\prime}=j^0=i(M\Pi_M+{\wt M}\Pi_{\wt M})
=2A^0_D(M^\dagger M\,+\,\widetilde M^\dagger \widetilde M),
\eeq
where the prime reminds us that we are working in the context
of the dual Lagrangian, and we have assumed that $\partial_0 M=0$. This charge density appears in the equation of motion for 
$A_D^0$
\beq\label{ADsource}
\partial_i{\rm Im}\,[\tau_D(F^{0i}_D-i\widetilde F^{0i}_D)]=-8\pi\,A^0_D(M^\dagger M\,+\,\widetilde M^\dagger \widetilde M)
\eeq
After substituting the dual BPS equation (\ref{BPSD}) and (\ref{j0}) into eq.(\ref{ADsource}), we obtain
$$
\partial_i{\rm Re}\,[\tau_D\,\eta\,\partial^ia_D]=-{4\pi\over\sqrt{2}}\,\nu_e^{\,\prime}
$$
or, using $\tau_D\partial^ia_D=-\pu ia$, 
\beq\label{asourceprime}
\nabla^2\,{\rm Re}\,(\eta\,a)=-{4\pi\over\sqrt{2}}\,\nu_e^{\,\prime}.
\eeq
Thus, away from a spherically symmetric source of total 
charge $n_e'=n_m=1$, we can adopt 
the same solution (\ref{Cmono}) found in the previous section.

As long as $r$ is large enough that $a_D\ne 0$, the
vev of $M$ is forced to vanish, and $\nu_e^{\,\prime}=0$ with it. 
But, no matter what the value of $a_\infty$, we eventually
reach a radius $r_0$ at which $a_D=0$. This is 
because along the ray of constant phase on which 
$a_D(r)$ takes its values, the range of $a_D$ is 
semi-infinite (since $a(r)$ also has semi-infinite range). 
At least for a large set of boundary conditions on $a_\infty$,
$a_D$ passes through 0. As $a_\infty$ approaches the value $a_0$
of $a$ at the monopole point, $a_{D\infty}$ approaches 0.
For $a_\infty$ on the ``other side'' of the monopole point,
one might expect that 0 is no longer in the range of $a_D$; however, the sign of $a_{D\infty}$ in eq.(\ref{Cmono}) flips, 
the semi-infinite ray of the range of $a_D$ now points in the 
opposite direction, and 0 is still included.  A numerical analysis of various values of $a_\infty$ confirms the conclusion that 
an $r_0$ for which $a_D(r_0)=0$ always exists. 

At this radius the potential terms for $M$ vanish 
and it becomes energetically
favorable for electric charge to be localized here. 
If for $r$ just below $r_0$, $a_D \ne 0$, then the
charge density 
will be a delta-function localized at radius $r_0$. 
If all the charge is localized at $r_0$, then
we get 
\beq\label{asource}
\nabla^2\,{\rm Re}\,(\eta\,a)=-\frac{1}{\sqrt{2}}\,\frac{\delta(r-r_0)}{r_0^2}.
\eeq
Then, for $r< r_0$, $a(r)$ will be constant, according to the BPS equation, and $a_D(r)$ will be identically zero. For the 
case $\theta=0$, the solution for $a(r)$ is
\beq\label{solution}
a(r)=\left\{\begin{array}{lll}
a_\infty+\frac{C}{\sqrt{2}\,r},&{\rm for~} r>r_0\\
&\\
a_\infty+\frac{C}{\sqrt{2}\,r_0},&{\rm for~} r\le r_0
\end{array}
\right.
\eeq
The discontinuity in the first derivative 
of $a(r)$ at $r_0$ is accounted for by the 
delta-function in eq.(\ref{asource}). 

\bigskip
We still need to check that our solution satisfies the 
equation of motion for $\bar a_D$, which is
\begin{eqnarray*}
&&\partial_{\mu}\tau_D\partial^{\mu}a_D+2i({\rm Im}\,\tau_D)\partial_\mu\partial^\mu a_D-\frac{1}{4}\bar \tau_D'F_{D\mu\nu}(F_D^{\mu\nu}+i\widetilde F_D^{\mu\nu})\\&&\qquad=-\,(4i)(4\pi)a_D(M^\dagger M\,+\,\widetilde M^\dagger \widetilde M)-{(4\pi)^2\over2}\,{\bar\tau_D'\over({\rm Im}\,\tau_D)^2}\,(M^\dagger M\,+\,\widetilde M^\dagger \widetilde M)^2
\end{eqnarray*}
This can be rewritten as 
\begin{eqnarray*}
&&2\re(\eta\nabla^2a)-2\bar\tau_D\, \re(\eta\nabla^2 a_D)\\
&&\qquad=
-{8\pi\over \sqrt2}
A_D^0(M^\dagger M\,+\,\widetilde M^\dagger \widetilde M)
+{(4\pi)^2\over2}\,{\eta\,\bar\tau_D'\over({\rm Im}\,\tau_D)^2}\,(M^\dagger M\,+\,\widetilde M^\dagger \widetilde M)^2
\end{eqnarray*}
where we have used 
the first BPS equation in (\ref{BPSD})
to replace $a_D$ by $A^0_D$ 
in the first term on the right side 
($i\sqrt2 \eta\,
a_D=A^0_D$ up to 
a constant, which we see 
should be set equal to zero
in order to satisfy this equation). 
The first terms on the left and right sides 
are equal by eq.(\ref{asourceprime}). 
The second term on the right side 
can be shown, by using
the BPS equations for $f_D$ and $D_D$,
to vanish identically, leaving
$$
{\rm Re}(\eta\nabla^2 a_D)=0,
$$
reproducing the result of eq.(\ref{harmaD}).

Note that we could now repeat the entire preceding analysis of this
section for a dyon of charge (1,1).  Like the monopole, its mass vanishes at a special point in moduli space.  There is a dual action 
${\cal L}_{D'}$ containing the dyon field, which becomes weakly coupled near to this point, in which the dyon appears electrically coupled to a dual gauge field.  The duality transformation which relates the original action to ${\cal L}_{D'}$ differs from the duality transformation employed above by a factor of $T$. From the BPS equations derived from ${\cal L}_{D'}$, we can derive a solutions with a very 
similar structure, including a spherical shell of dyonic charge, 
for the dyon. 
However, this procedure only works for the charges (1,0) and (1,1),
since these are the only states that become massless at some point.

\bigskip

Our understanding of the structure of monopoles and dyons in the U(1) effective theory now allows us to complete the proof of the central charge formula.

Let us begin by considering $Z$ in the presence of a pure magnetic monopole $(n_e=0)$. Using the equation of motion $\partial_i\Pi_{A_i}=0$, the central charge (\ref{line1}-\ref{line2}) reduces to
\beq\label{Zmono}
Z=a_{D\infty}\,n_m+\frac{1}{4\pi}\,\int\,d^3x\,\left(a_D\,\partial_i\widetilde F^{0i}\right),
\eeq

In showing that the second term on the right-hand side of (\ref{Zmono}) vanishes, giving $Z=a_{D\infty}\,n_m$, it would be incorrect to invoke the Bianchi identity $\partial_i\widetilde F^{0i}=0$ (as in \cite{Iorio2}), because by (\ref{integers}) we would then have $n_m=0$, so $Z=0$. The present calculation is different from the one for the central charge in the \nn  super-Yang-Mills theory with gauge group SU(2) \cite{Alv,Wei,Wolf} where the Bianchi identity for the non-Abelian gauge field can be used because it is not directly related to the magnetic charge\footnote{In the SU(2) gauge theory, the electric and magnetic charges are identified as
$$
Q_e=-{1\over\phi_\infty g}\int\,d^3x\,\partial_ i(\phi^aF^{a0i})\quad Q_m=-{1\over\phi_\infty g}\int\,d^3x\,\partial_i(\phi^a\widetilde F^{a0i})
$$
where $a=1,2,3$ is the gauge index that labels the adjoint representation of SU(2), $F^{a0i}$ is the non-Abelian gauge field, and $\phi_\infty$ is the asymptotic value of the scalar field triplet, $(\phi^a\phi^a)^{1\over 2}$ (in the Higgs vacuum).}, and is always satisfied.

Instead, as we have shown in this section, the magnetic charge is
located only where $a_D(r)=0$, i.e. on a sphere of radius $r_0$. Then,
$$
\partial_i\widetilde F^{i0}={n_m\over r^2}\,\delta(r-r_0),
$$ 
so the integral in (\ref{Zmono}) vanishes and we end up with $$
Z=a_{D\infty}n_m.
$$

Using duality, we can now generalize to any dyon. In this case $\partial_i\Pi_{A_i}\ne0$, and the electric and magnetic charges are distributed over a sphere of radius $r_0$, defined by the duality-invariant quantity $Z_0=Z(r_0)=a_0n_e+a_{D0}n_m=0$, where the dyon appears locally to be  massless (in the sense of a ``local mass'' defined as $\sqrt2\left|Z(r)\right|$ ). Since the terms (\ref{line2}) are equal to $-Z_0$, we obtain
$$
Z=a_{\infty}\,n_e+a_{D\infty}\,n_m
$$

%As already noted, $M$ vanishes when $a_D\ne 0$, or, likewise,
%when $ \tau_D \ne \infty$. 
%So we can equate the first term on the right side of 
%this equation to 0.  Also, 

\bigskip

\section{Dyons and Sources}

\bigskip
In the previous section we have seen how the solution (\ref{solution}) emerges in the dual description of the theory, which is the natural framework for describing monopoles. We have coupled the dual  \nn  vector multiplet to an \nn  hypermultiplet representing the source of the former. On the other hand, the original Lagrangian (2) describing the  vector multiplet cannot be coupled in a similar way to a source hypermultiplet, because monopoles do not appear as elementary degrees of freedom in that description of the theory and cannot couple locally to the gauge field. We can still use the Lagrangian (2), but only at the expense of describing the monopole as a classical source. This is equivalent to treating the electric and magnetic fields of the source as background fields. One of the most important consequences of this approach is that it will provide the solution of the problem of the extra term in the supersymmetry variation (\ref{deltaL}) of the Lagrangian. Thus, we complete the description of the sources of the theory at both weak and strong coupling.

In section 4 we have already written down the equations of motion with source terms on the right-hand side. The Lagrangian (2), as it stands, cannot reproduce those sources because it only describes the free-field vector multiplet. In order to derive the equations of motion in the presence of electric and magnetic sources, we have to add a source term to the original Lagrangian. As we will check explicitly below, that the addition should be
\br
{\cal L}_{source}&=&{\rm Im}\,\left[\sqrt{2}\,\eta\,(a\,\nu_e+a_D\,\nu_m)\,-\,iA_\mu\,j_e^\mu\right]\nonumber\\
&=&{\rm Im}\,\left[\sqrt{2}\,\eta\,(a\,\nu_e+a_D\,\nu_m)\,-\,iA_0\,\nu_e\right].\label{Lsource}
\er
for static fields and sources. 
%This expression agrees with the fact that only the electric charges can couple locally to the U(1) gauge field. 
%Throughout this section it should be understood that $n_m=\pm 1$, while $n_e$ can be any integer, since these are the only spherically symmetric dyons, and they are obtained from the simple monopole solution by T-duality, as seen in section 4. 
In the presence of (\ref{Lsource}), the equation of motion for the scalar field $\bar a$ becomes
\begin{equation}\label{neweqa}
\partial_{\mu}\tau\partial^{\mu}a+2i({\rm Im}\,\tau)\partial_\mu\partial^\mu a-\frac{1}{4}\bar\tau'F_{\mu\nu}(F^{\mu\nu}+i\widetilde F^{\mu\nu})=-\bar\eta\,\sqrt{2}(4\pi)(\nu_e+\bar\tau\nu_m)
\end{equation}
where the fermionic terms have been again omitted since we are interested in the equation satisfied by a BPS configuration. Notice that (\ref{neweqa}), just as its source-free counterpart (\ref{eqfora}), has the property of being invariant only under T-duality. Using the BPS equations (\ref{BPS}-\ref{bps}) and following the steps that led to eq. (\ref{harma}), we obtain 
\begin{equation}\label{newharma}
{\rm Re}(\eta\nabla^2a)=\frac{1}{\sqrt{2}}\,\frac{{\rm Im}(n_e+\bar\tau_0n_m)}{{\rm Im}\,\tau_0}\,\frac{\delta(r-r_0)}{r_0^2}=-\frac{4\pi}{\sqrt{2}}\,\nu_m
\end{equation}
Let us consider again the equation of motion for the scalar field in the presence of the source, written as
$$
{\rm Re}(\eta\partial_i\tau\partial^i a)-i\eta({\rm Im}\,\tau)\nabla^2a=-\frac{4\pi}{\sqrt{2}}\,(\nu_e+\bar\tau\nu_m)
$$
Adding it to its complex conjugate, we get
\begin{equation}\label{complicatedeq}
{\rm Re}(\eta\partial_i\tau\partial^i a)+({\rm Im}\,\tau)\,{\rm Im}(\eta\,\nabla^2a)=-\frac{4\pi}{\sqrt{2}}\,\left[\nu_e+({\rm Re}\,\tau)\nu_m\right].
\end{equation}

The equation of motion for the gauge field $A_\mu$ in the presence of the source (\ref{Lsource}) is,
$$
-\frac{1}{4\pi}\,\partial_\mu\,{\rm Im}\,\left[\tau(F^{\mu\nu}-i\widetilde F^{\mu\nu})\,+\,\sqrt{2}\,\tau'(\lambda\,\sigma^{\mu\nu}\,\psi)\right]=\frac{\partial\cal L}{\partial A_\nu},\quad {\rm for~}\nu=0,1,2,3.
$$
Only the equation for $\nu=0$ has a non-zero right-hand side. Using (\ref{canmom}), it can be expressed as
$$
\partial_i\,\Pi_{A_i}=-\frac{\partial\cal L}{\partial A^0}=\nu_e
$$
For a bosonic BPS state, the above equation reduces to
$$
\partial_i{\rm Im}[\tau(F^{0i}-i\widetilde F^{0i})]=-4\pi\nu_e
$$
Substituting the BPS equation (\ref{BPS}), we obtain
\beq\label{newharmaD}
{\rm Re}\left(\eta\nabla^2 a_D\right)=\frac{4\pi}{\sqrt{2}}\,\nu_e
\eeq
Notice that the signs in eqs. (\ref{newharma}) and (\ref{newharmaD}) agree with the fact that the latter is the S-transformed of the former. Indeed, under S-duality (see Appendix F), $a\to a_D$, $a_D\to-a$ and $(n_m,n_e)\to(-n_e,n_m)$. We can also write this equation as
$$
{\rm Re}(\eta\partial_i\tau\partial^i a)-{\rm Re}\left[\tau\nabla^2(\eta a)\right]=-\frac{4\pi}{\sqrt{2}}\,\nu_e
$$
and it is easy to see that it is identical to eq.(\ref{complicatedeq}).

\bigskip
We are now in a position to elucidate the mystery of the additional term which occurred in the \none  supersymmetry variation (\ref{deltaL}) of the Lagrangian (2),
\beq\label{Bianchi}
\frac{1}{4\pi}\,{\rm Im}\,[2\,\bar\tau(\xi\,\sigma^\nu\bar\lambda)(\partial^\mu\widetilde F_{\mu\nu})]
\eeq
If the Bianchi identity can be applied (i.e. if $n_m=0$), then the Lagrangian is invariant under \none  supersymmetry transformations\footnote{As mentioned earlier, even in the most basic case of the Lagrangian for the \none vector superfield \cite{WB}, in order to obtain the variation of the Lagrangian as a total divergence, the Bianchi identity must be imposed \cite{Wei}.}. The Lagrangian (2) is inadequate to describe monopoles and dyons, and it can do so only in terms of a classical source, as seen in the present section. In contrast, the dual Lagrangian coupled to a source hypermultiplet incorporates monopoles as elementary degrees of freedom, as we have seen in the previous section. However, the fact remains that in the presence of a monopole ($n_m\ne 0$), when we allow a non-zero $a_Dn_m$ term in $Z$, we cannot set $\partial^\mu\widetilde F_{\mu\nu}$ equal to zero, because this is precisely where $n_m$ comes from. Fortunately, by including ${\cal L}_{source}$ in the Lagrangian (2), we obtain a totally consistent picture in which the \nn  supersymmetry variation of ${\cal L}_{source}$ cancels exactly the \nn  version of (\ref{Bianchi}), without being necessary to impose the Bianchi identity. This shows that the supersymmetry currents obtained in section 2 are correct in all cases, without anything being arbitrarily discarded. The following proof of invariance uses the relation between the two supersymmetry parameters $\xi^{(1)\alpha}$ and $\bar\xi^{(2)}_{\dot\beta}$, given in section 4, which is the key to the physical meaning of the cancellation of the term (\ref{Bianchi}) by the variation of the source term representing a BPS state. In general, a monopole breaks supersymmetry (through a term (\ref{Bianchi}) that exists outside a total divergence), but for each BPS configuration a certain linear combination of generators is possible such that the fields of the \nn  multiplet are left invariant by half of the generators. We will now show that this invariance is also manifest at the level of the Lagrangian, when the required linear combination is realized.

Using
$$
n_m={1\over 4\pi}\int d\,^3\,x\,\pd i\widetilde F^{i0}=\int d\,^3\,x\,\nu_m
$$
the term (\ref{Bianchi}) becomes, for a static distribution of magnetic charge,
$$
2\,\nu_m\,{\rm Im}\,[\bar\tau(\xi^{(1)}\,\sigma^0\bar\lambda)]
$$
This is the result of one of the \none variations of (2). The corresponding (\ref{Bianchi}) term for the \nn -supersymmetry variation of the Lagrangian is
\br
&&2\,\nu_m\,{\rm Im}\,[\bar\tau(\xi^{(1)}\,\sigma^0\bar\lambda+\xi^{(2)}\,\sigma^0\bar\psi)]\nonumber\\
&&=-2\,\nu_m\,{\rm Im}\,[\tau(\lambda\,\sigma^0\bar\xi^{(1)}+\psi\,\sigma^0\bar\xi^{(2)})]\nonumber\\
&&=-2\,\nu_m\,{\rm Im}\,[\tau(-\eta\,\xi^{(2)}\lambda+\eta\,\xi^{(1)}\psi)]\label{Bi}
\er
where in the last step we have used (\ref{xi}). Consider now the \nn  variation of the source term ${\cal L}_{source}$, written as
$$
\delta{\cal L}_{source}=\left.\delta{\cal L}_{source}\right|_{\nu_e=0}+\left.\delta{\cal L}_{source}\right|_{\nu_m=0}
$$
where 
\beq\label{-Bi}
\left.\delta{\cal L}_{source}\right|_{\nu_e=0}=\sqrt{2}\,\nu_m\,{\rm Im}\,(\eta\,\tau\delta a)
\eeq
and
\beq\label{el}
\left.\delta{\cal L}_{source}\right|_{\nu_m=0}=-\nu_e\left[\delta A^0+{i\over\sqrt{2}}\,(\eta\,\delta a-\bar\eta\,\delta\bar a)\right]
\eeq
The reason for this splitting of $\delta{\cal L}_{source}$ is that the piece which cancels (\ref{Bi}) is (\ref{-Bi}), while (\ref{el}) vanishes. Indeed, 
$$
\left.\delta{\cal L}_{source}\right|_{\nu_e=0}=2\,\nu_m\,{\rm Im}\,[\eta\,\tau(\xi^{(1)}\psi-\xi^{(2)}\lambda)]
$$
cancels (\ref{Bi}), and
\brs
\left.\delta{\cal L}_{source}\right|_{\nu_m=0}&=&-i\,\nu_e\left(\bar\xi^{(1)}\bar\sigma^0\lambda-\bar\lambda\bar\sigma^0\xi^{(1)}+\bar\xi^{(2)}\bar\sigma^0\psi-\bar\psi\bar\sigma^0\xi^{(2)}\right.\\
&&\hspace{3cm}\left.+\eta\,\xi^{(1)}\psi-\eta\,\xi^{(2)}\lambda-\bar\eta\,\bar\xi^{(1)}\bar\psi+\bar\eta\,\bar\xi^{(2)}\bar\lambda\right)   
\ers
Using (\ref{xi}) in the first four terms of the right-hand side of the above equation, we obtain
$$
\left.\delta{\cal L}_{source}\right|_{\nu_m=0}=0
$$
In a manner of speaking, the Lagrangian (2) with ${\cal L}_{source}$ added is not supersymmetric at the \none level in the presence of monopoles or dyons, but becomes supersymmetric at the \nn  level if the monopoles or the dyons are BPS states. Superficially, it may seem that supersymmetry is a special type of symmetry from the point of view of Noether's theorem, since an additional constraint\footnote{In the absence of sources, the constraint is the Bianchi identity; in the presence of sources, the constraint is that the source is a BPS state.} must be imposed in order to realize this symmetry on the Lagrangian. However the situation is perfectly analogous to the fact that gauge invariance is broken in the presence of a prescribed external source, but is restored if the source is included in the Lagrangian, as it will be seen shortly.

\bigskip
Finally, we can reproduce the BPS mass formula (\ref{M3}) by calculating directly the energy of a BPS field configuration. 
%[ref\{Physica {\bf6}, 887 (1939)\}], 
Constructing the Belinfante symmetric energy-momentum tensor $T^{\mu\nu}$ from the original Lagrangian (2), we find
\br
T^{00}&=&{1\over 4\pi}\,{\rm Im}\left[\tau{\bf\nabla}a\cdot{\bf\nabla}\bar a-{\tau\over 2}\,(F_{0i}+i\widetilde F_{0i})(F^{0i}-i\widetilde F^{0i})\right]-A^0(\pd i\Pi_{A_i})\nonumber\\
&=&{1\over g^2}\,{\bf\nabla}a\cdot{\bf\nabla}\bar a+{1\over 2}(E^2+B^2)-A^0(\pd i\Pi_{A_i})\label{T00}
\er
If $n_e=0$, as it would follow from the free-field Lagrangian (2), the last term in (\ref{T00}) vanishes. However, if $n_e\ne0$, that last term appears to violate gauge invariance. This situation is familiar from classical electrodynamics in the case of an open system with a prescribed external source. The solution of this problem is to include the source in the system, thereby restoring gauge invariance. If we include ${\cal L}_{source}$ in (2), then we have to subtract it from the right-hand side of (\ref{T00}),
\brs
T^{00}&=&{1\over g^2}\,{\bf\nabla}a\cdot{\bf\nabla}\bar a+{1\over 2}(E^2+B^2)-A^0(\pd i\Pi_{A_i})\nonumber\\
&&-{\rm Im}\,\left[\sqrt{2}\,\eta\,(a\,\nu_e+a_D\,\nu_m)\,-\,iA^0\,\nu_e\right]
\ers
We notice that the $A^0$ terms disappear from $T^{00}$, restoring gauge invariance. When integrating over all space to obtain $M$, we recognize (\ref{line2again}) in the remaining contribution of the source term, giving $\sqrt{2}\,\left|Z_0\right|$, which is equal to zero, as discussed at the end of section 5. Hence, we end up with the following expression for the mass of a bosonic field configuration
\beq
M=\int\,d\,^3\,x\,\left[{1\over g^2}\,{\bf\nabla}a\cdot{\bf\nabla}\bar a+{1\over 2}(E^2+B^2)\right],\label{mass}
\eeq
In particular, for a BPS state, substituting (\ref{BPS4}) into (\ref{mass}), we get
\beq
M=2\int\,d\,^3\,x\,{1\over g^2}\,{\bf\nabla}a\cdot{\bf\nabla}\bar a\label{vir}
\eeq
Eq. (\ref{vir}) is the same as (\ref{Z2}) inserted in the mass bound (\ref{M3}). Comparison of (\ref{mass}) and (\ref{vir}) shows the very interesting fact that, in a BPS field configuration, exactly half of the energy is stored in the scalar field and the other half is stored in the electromagnetic field. Therefore, the BPS mass formula can also be written as
\beq\label{eandm}
M=\int\,d\,^3\,x\,(E^2+B^2)
\eeq

\acknowledgments

We thank Philip Argyres, Adel Awad, and Nathan Berkovits
for useful discussions. We are grateful to 
Robert Jaffe and the MIT Center for Theoretical
Physics for hospitality while portions of this work were 
completed.
IP thanks the University of Kentucky for a Graduate 
Research Fellowship. 
This work was supported by NSF Grants PHY-9722147  and 
PHY-0071312 and DOE Grant No. DE-FG01-00ER45832.

\vspace{2cm}

\appendix

\bigskip
\section{Notations and Conventions}

\bigskip
\subsection{Metric, Pauli Matrices, Spinors}
We use a Minkowski metric with signature $(+---)$. The conventions for supersymmetry are those of \cite{WB}, with a few exceptions that follow from our choice of metric and will be noted below. Our conventions are then the same as in \cite{BL}, excepting the fact that we take $\varepsilon^{0123}=+1$, as in \cite{WB}. To be specific, with our choice of metric, the right-hand sides of several relations from \cite{WB}, Appendix B, will have an opposite sign\footnote{The relations from \cite{WB}, Appendix B, for which the right-hand side changes sign are (B.4), (B.5), (B.11), (B.14), (B.19). The sign also changes for the first term on the right-hand side of the second relation (B.9).}. Another difference is that with the $(+---)$ metric, we must define $\sigma^0=\bar\sigma^0=1_{2\times 2}$ (which will be our choice), while $(-+++)$ require $\sigma^0=\bar\sigma^0=-1_{2\times 2}$ (as in \cite{WB}), and we cannot mix the signs between these two conventions because otherwise a relation that has a form independent of the signature of the metric, such as (B.8) from \cite{WB}, would not be consistent.

\bigskip
\subsection{Superfields}
The \none chiral superfield in the Lagrangian (1) has the following field content,
\brs
\Phi&=&a+\sqrt{2}\theta\psi+i(\theta\sigma^\mu\bar\theta)\partial_\mu a -\frac{i}{\sqrt{2}}(\theta\theta)(\partial_\mu\psi\sigma^\mu\bar\theta)\\&&-\frac{1}{4}(\theta\theta)(\bar\theta\bar\theta)\partial_\mu\partial^\mu a+(\theta\theta)f
\ers
(we denote the scalar component of this superfield by $a$, as in \cite{SW1}). The corresponding antichiral superfield and dual superfield will be denoted by $\bar\Phi$ and $\Phi_D$, respectively. In order to establish notation, we also give the field content of the superfield ${\cal M}$ of the dual Lagrangian ${\cal L}_D$:
\brs
{\cal M}&=&M+\sqrt{2}\theta\psi_M+i(\theta\sigma^\mu\bar\theta)\partial_\mu M -\frac{i}{\sqrt{2}}(\theta\theta)(\partial_\mu\psi_M\sigma^\mu\bar\theta)\\&&-\frac{1}{4}(\theta\theta)(\bar\theta\bar\theta)\partial_\mu\partial^\mu M+(\theta\theta)f_M,
\ers
with a similar notation for $\widetilde{\cal M}$. The vector superfield in the Wess-Zumino gauge is
$$
V_{WZ}=\theta\sigma^\mu\bar\theta A_\mu+i(\theta\theta)\bar\theta\bar\lambda-i(\bar\theta\bar\theta)\theta\lambda+{1\over2}\,(\theta\theta)(\bar\theta\bar\theta)D
$$
Notice that this expresion is the one from \cite{BL}, and differs from the one in \cite{WB}, Eq.(6.6), by the sign of the first term. The reason for this sign difference is that, in going from (4.9) to (6.2) in \cite{WB}, the sign of the term written there as $\theta\sigma^m\bar\theta v_m$ is conventionally changed from + to -\,. We choose not to make this change, hence we follow \cite{BL}. The consequences of this difference are: 

1) in the field-strength superfield $W_\alpha$, the term $i(\sigma^{\mu\nu}\theta)_\alpha F_{\mu\nu}$ is now preceded by a plus sign, while in \cite{WB} it has a minus sign,  

2) in the supersymmetry variation $\delta\lambda$, the term in $F_{\mu\nu}$ is now preceded by a minus sign, instead of a plus sign as in \cite{WB}, and

3) in the Lagrangian (2), the term in $(\lambda\sigma^{\mu\nu}\psi)F_{\mu\nu}$ has a minus sign in our convention, instead of a plus sign.

Obviously, with either convention, the physical results are the same.

\bigskip
\section{Derivation of the U(1) Effective Lagrangian}
The first term in the Lagrangian (1) is the coefficient of the $(\theta\theta)(\bar\theta\bar\theta)$ term (known as the D-term) in the expansion of the K\"{a}hler potential,
$$
\int d^2\theta\int d^2\bar\theta\,K(\Phi,\bar\Phi),
$$
where $\Phi$ is the chiral superfield given above. Denoting $\Phi\equiv a+\Delta\Phi$ and expanding the K\"{a}hler potential in a Taylor series, we have
\br
&&\int d^2\theta\int d^2\bar\theta\,K(\Phi,\bar\Phi)=-\left({1\over 4}\,{\p K\over\p a}\,\pd\mu\pu\mu a+{1\over 4}\,{\pu 2K\over\p a^2}\,\pd\mu a\pu\mu a\right.\nonumber\\
&&\left.+{i\over 2}\,{\pu 2K\over\p a\p\bar a}\,\psi\sigma^\mu\pd\mu\bar\psi+{i\over 2}\,{\pu 3K\over\p a\p\bar a^2}(\pd\mu\bar a)(\psi\sigma^\mu\bar\psi)+{1\over 2}\,{\pu 3K\over\p a^2\p\bar a}(\psi\psi)\bar f\right)\,+\,c.c.\nonumber\\
&&+{\pu 2K\over\p a\p\bar a}\left({1\over 2}\,\pd\mu a\pu\mu\bar a+f\bar f\right)+{1\over 4}\,{\pu 4K\over\p a^2\p\bar a^2}(\psi\psi)(\bar\psi\bar\psi)\label{kahler}
\er
The terms of (\ref{kahler}) involving only $a$ and $\bar a$ can be combined into
$$
{\pu 2K\over\p a\p\bar a}\,\pd\mu a\pu\mu\bar a\,-\,\pd\mu\left({1\over 4}\,{\p K\over\p a}\,\pu\mu a+c.c.\right)
$$
and we drop the total divergence.

In terms of the holomorphic prepotential ${\cal F}(\Phi)$ which depends only on the superfield $\Phi$, the K\"{a}hler potential of the present theory can be expressed as
$$
K(\Phi,\bar\Phi)={\rm Im}\left({\p{\cal F}\over\p\Phi}\,\bar\Phi\right)
$$
For the K\"{a}hler metric we have
$$
g_{a\bar a}={\rm Im}\,\tau(a)={\rm Im}\,{\pu 2{\cal F}\over\p a^2}={\pu 2K\over\p a\p\bar a}
$$
Also,
$$
{\pu 3K\over\p a^2\p\bar a}={1\over 2i}\,{\p\tau\over\p a},\quad{\pu 3K\over\p a\p\bar a^2}=-{1\over 2i}\,{\p\bar\tau\over\p\bar a},\quad{\pu 4K\over\p a^2\p\bar a^2}=0.
$$
The terms
$$
-\left({i\over 2}\,{\pu 2K\over\p a\p\bar a}\,\psi\sigma^\mu\pd\mu\bar\psi+{i\over 2}\,{\pu 3K\over\p a\p\bar a^2}(\pd\mu\bar a)(\psi\sigma^\mu\bar\psi)\right)+c.c.
$$
of (\ref{kahler}) reduce to ${\rm Im}\,(-i\tau\psi\sigma^\mu\pd\mu\bar\psi)$, where we have dropped a total divergence. Then (\ref{kahler}) becomes
\br
&&\int d^2\theta\int d^2\bar\theta\,K(\Phi,\bar\Phi)={\rm Im}\int d^2\theta\int d^2\bar\theta\,{\p{\cal F}\over\p\Phi}\,\bar\Phi\nonumber\\
&&={\rm Im}\left[\tau(a)\left(\pd\mu a\pu\mu\bar a-i\psi\sigma^\mu\pd\mu\bar\psi+f\bar f\right)-{1\over 2}\,\tau'(\psi\psi)\bar f\right],\label{firstterm}
\er
where $\tau'\equiv\p\tau/\p a$.

Let us now turn our attention to the second term in the Lagrangian (1) which is the coefficient of the $(\theta\theta)$ term (known as the F-term) in the expansion of $\tau(\Phi)W^\alpha W_\alpha$, where
$$
\tau(\Phi)={\pu 2{\cal F}\over\p\Phi^2}=\tau(a)+{\p\tau\over\p a}\,\Delta\Phi+{1\over 2}\,{\pu 2\tau\over\p a^2}\,(\Delta\Phi)^2
$$
(all higher-order terms vanish). We have
\br
W^\alpha W_\alpha&=&-\lambda^\alpha\lambda_\alpha-2i\lambda^\alpha\left[\delta_\alpha\,^\beta D+i(\sigma^{\mu\nu})_\alpha\,^\beta F_{\mu\nu}\right]\theta_\beta\nonumber\\
&&+(\theta\theta)\left[D^2-2i\lambda^\alpha(\sigma^\mu)_{\alpha\dot\alpha}\pd\mu\bar\lambda^{\dot\alpha}-{1\over 2}\,F_{\mu\nu}(F^{\mu\nu}-i\widetilde F^{\mu\nu})\right]\label{WW}
\er
It should be noted that in this formulation the component fields depend on the {\em superspace} coordinate $y^\mu=x^\mu+i\theta\sigma^\mu\bar\theta$. We obtain
\br
\int d^2\theta\,\tau(\Phi)W^\alpha W_\alpha&=&\tau\left[D^2-2i\lambda\sigma^\mu\pd\mu\bar\lambda-{1\over 2}\,F_{\mu\nu}(F^{\mu\nu}-i\widetilde F^{\mu\nu})\right]\nonumber\\
&&+\tau'\left[i\sqrt{2}\,D(\psi\lambda)-\sqrt{2}(\lambda\sigma^{\mu\nu}\psi)F_{\mu\nu}-(\lambda\lambda)f\right]\nonumber\\
&&+{1\over 2}\,\tau''(\lambda\lambda)(\psi\psi)\label{secondterm}
\er
Then, using (\ref{firstterm}) and (\ref{secondterm}) in (1), we get the effective Lagrangian (2) in terms of component fields.

\bigskip
\section{Constraints and Dirac Brackets}
All commutators and anticommutators that appear in the main text are Dirac brackets and are denoted simply by [ , ] or \{ , \}. The more familiar canonical (Poisson) brackets will be indicated by a subscript 'P'. The constraints referred to below are relations between canonical momenta and fields. For example, the fact that
$$
(\Pi_\psi)_\alpha=-\frac{\bar\tau}{8\pi}(\sigma^0)_{\alpha\dot\beta}\bar\psi^{\dot\beta},
$$
gives the constraint
$$
(\theta_1)_\alpha=(\Pi_\psi)_\alpha+\frac{\bar\tau}{8\pi}(\sigma^0)_{\alpha\dot\beta}\bar\psi^{\dot\beta}=0.
$$
When such constraints are present in a system, the canonical quantization proceeds with Dirac brackets (to be defined below), rather than with canonical brackets. For more details, we refer the reader to \cite{HandT}. Below we list the second class constraints for the Lagrangian (2):
\brs
(\theta_1)_\alpha&=&(\Pi_\psi)_\alpha+\frac{\bar\tau}{8\pi}(\sigma^0)_{\alpha\dot\beta}\bar\psi^{\dot\beta},\\
(\theta_2)_{\dot\alpha}&=&(\Pi_{\bar\psi})_{\dot\alpha}+\frac{\tau}{8\pi}\psi^{\beta}(\sigma^0)_{\beta\dot\alpha},\\
\theta_3&=&f+{i\over4\im\tau}\,(\tau'\psi\psi-\bar\tau'\bar\lambda\bar\lambda),\\
\theta_4&=&D+{1\over2\sqrt2}\,{1\over\im\tau}\,(\tau'\psi\lambda-\bar\tau'\bar\psi\bar\lambda),\\
\theta_5&=&\bar f-{i\over4\im\tau}\,(\bar\tau'\bar\psi\bar\psi-\tau'\lambda\lambda),\\
\theta_6&=&\Pi_f,\quad\theta_7\;=\;\Pi_D,\quad\theta_8\;=\;\Pi_{\bar f},\\
(\theta_9)_\alpha&=&(\Pi_\lambda)_\alpha+\frac{\bar\tau}{8\pi}(\sigma^0)_{\alpha\dot\beta}\bar\lambda^{\dot\beta},\\
(\theta_{10})_\alpha&=&(\Pi_{\bar\lambda})_{\dot\alpha}+\frac{\tau}{8\pi}\lambda^{\beta}(\sigma^0)_{\beta\dot\alpha}.
\ers
These constraints are used to construct a nonsingular matrix $C$, of elements
$$
c_{ij}=[\theta_i,\theta_j]_P
$$
In order to compute the inverse $C^{-1}$ of this matrix, we first have to expand it as:
$$
C(q^i,\psi^\alpha)=c_0(q^i)+C_\alpha(q^i)\psi^\alpha,
$$
where $q^i$ and $\psi^\alpha$ are bosonic and fermionic variables, respectively. For the \nn  vector multiplet, the above expression becomes
$$
C=C_0+C_1\psi_\alpha+C_2\lambda_\alpha+C_3\bar\psi_{\dot\alpha}+C_4\bar\lambda_{\dot\alpha}.
$$
The inverse matrix, of elements $c^{jk}$ (with $c_{ij}c^{jk}=\delta_i\,^k$), is given by
$$
C^{-1}\equiv B=B_0+B_\alpha\psi^\alpha,
$$
where
$$
B_0=C_0^{-1},\qquad B_\alpha=-C_0^{-1}C_\alpha C_0^{-1}.
$$
The large matrix $C$ is a block matrix, so only parts of it may be inverted, as needed. For example, the calculation of the central charge (Appendix E) does not involve the constraints $\theta_3,\ldots,\theta_8$ which can be left out of the matrix $C$. Then, the Dirac bracket of any two fields and/or momenta $\phi$ and $\chi$ is defined in terms of the inverse matrix and canonical brackets as follows
\beq\label{defDirac}
[\phi,\chi]=[\phi,\chi]_P-[\phi,\theta_j]_Pc^{ji}[\theta_i,\chi]_P.
\eeq
Below we list some of the Dirac brackets used in the calculation of the central charge:
\brs
\left[\Pi_a,\Pi_{\bar a}\right]&=&-\delta^3\frac{1}{16\pi}\,\frac{\tau'\bar\tau'}{{\rm Im}\,\tau}\,(\psi\,\sigma^o\bar\psi+\lambda\,\sigma^o\bar\lambda)\\
\left[\Pi_a,\psi_\gamma\right]&=&\delta^3\frac{\tau'}{2\,{\rm Im}\,\tau}\,\psi_\gamma,\qquad \left[\Pi_a,\lambda_\gamma\right]\;=\;\delta^3\frac{\tau'}{2\,{\rm Im}\,\tau}\,\lambda_\gamma\\
\left[\Pi_a,\bar\psi_{\dot\alpha}\right]&=&\left[\Pi_a,\bar\lambda_{\dot\alpha}\right]\;=\;0,\quad\left[\Pi_{\bar a},\lambda_{\alpha}\right]\;=\;0,\quad\left[\Pi_{\bar a},\psi_{\alpha}\right]\;=\;0\\
\left\{\psi_\alpha,\bar\psi^{\dot\alpha}\right\}&=&\left\{\psi_\alpha,\bar\psi^{\dot\alpha}\right\}\;=\;-\delta^3\frac{4\pi}{{\rm Im}\,\tau}\,(\sigma^0)_{\alpha\dot\alpha}\\
\left\{\psi_\alpha,{\Pi_\psi}_\beta\right\}&=&\left\{\lambda_\alpha,{\Pi_\lambda}_\beta\right\}\;=\;\delta^3\frac{\bar\tau}{2\,{\rm Im}\,\tau}\,\varepsilon_{\alpha\beta}
\ers
where $\delta^3$ is an abbreviation for $\delta^{(3)}({\bf x}-{\bf y})$.

For example, let us calculate the Dirac bracket $\{\bar\psi^{\dot\alpha},\psi_\alpha\}$. The only matrix element contributing to it is $c^{21}$. We have
$$
(c_{12})_{\alpha\dot\alpha}=\{(\theta_1)_\alpha,(\theta_2)_{\dot\alpha}\}_P=\delta^3\,{{\rm Im}\,\tau\over4\pi}\,(\sigma^0)_{\alpha\dot\alpha},
$$
so
\brs
\{\bar\psi^{\dot\alpha},\psi_\alpha\}&=&-\{\bar\psi^{\dot\alpha},(\theta_2)_{\dot\beta}\}_P(c^{21})^{\dot\beta\beta}\{(\theta_1)_\beta,\psi_\alpha\}_P\\
&=&-\left(i\delta^3\delta_{\dot\beta}^{\dot\alpha}\right)\left[(\delta^3)^{-1}\,{4\pi\over{\rm Im}\,\tau}\,(\bar\sigma^0)^{\dot\beta\beta}\right]\left(i\delta^3\varepsilon_{\beta\alpha}\right)\;=\;\delta^3\,{4\pi\over{\rm Im}\,\tau}\,(\bar\sigma^0)^{\dot\alpha\beta}\varepsilon_{\beta\alpha}
\ers
As a consistency check, we may use this result to compute $\left\{\Pi_{\psi\beta},\psi_\alpha\right\}$ in two different ways: directly from the definition of the Dirac bracket (and in this case the first term on the right-hand side of (\ref{defDirac}) is non-zero), or by writing the definition of $\Pi_{\psi\beta}$ as a canonical momentum and using $\{\bar\psi^{\dot\alpha},\psi_\alpha\}$. The result will obviously be the same. In fact, the Dirac constraint formalism allows us to interchange freely the fields and their momenta, if their constraint relation is properly taken into account, allowing us to perform calculations, such as the one for the central charge, in terms of either a combination of fields and momenta or just fields alone. (Incidentally, we did not find it necessary to adopt, for raising and lowering the spinor indices of momenta conjugated to fermion fields, a rule opposite to the one for the fields themselves, as in \cite{Iorio2}.)

\bigskip
\section{\none  Supersymmetry Variation of the U(1) Effective Lagrangian}

\bigskip
\subsection{Infinitesimal Supersymmetry Transformations}
The variations of the component fields of the Lagrangian (2) under an \none transformation with parameter $\xi$ are
\brs
\delta a&=&\sqrt{2}\,\xi\psi\quad({\rm so~}\delta\tau=\sqrt{2}\,\tau'\,\xi\psi),\qquad\delta\bar a\,\;=\,\;\sqrt{2}\,\bar\xi\bar\psi,\\
\delta\psi&=&i\sqrt{2}\,\sigma^\mu\,\bar\xi\,\partial_\mu a\,+\,\sqrt{2}\,\xi f,\\
\delta\bar\psi&=&i\sqrt{2}\,\bar\sigma^\mu\,\xi\,\partial_\mu \bar a\,+\,\sqrt{2}\,\bar\xi\bar f,\\
\delta f&=&i\sqrt{2}\,\bar\xi\,\bar\sigma^\mu\,\partial_\mu \psi,\qquad \delta\bar f\,\;=\,\;i\sqrt{2}\,\xi\,\sigma^\mu\,\partial_\mu \bar\psi,
\ers
\brs
\delta A^\mu&=&i(\bar\xi\,\bar\sigma^\mu\lambda\,-\,\bar\lambda\,\bar\sigma^\mu\xi),\quad{\rm or}\\
\delta F^{\mu\nu}&=&i[(\xi\,\sigma^\nu \partial^\mu\bar\lambda\,+\,\bar\xi\,\bar\sigma^\nu\partial^\mu\lambda)\,-\,(\nu<\!\!\!--\!\!\!>\mu)],\\
\delta\lambda&=&-\frac{1}{2}\,\sigma^\mu\bar\sigma^\nu\xi F_{\mu\nu}\,+\,i\xi D,\\
\delta\bar\lambda&=&\frac{1}{2}\,\bar\xi\,\bar\sigma^\mu\sigma^\nu F_{\mu\nu}\,-\,i\bar\xi D,\\
\delta D&=&\partial_\mu(\bar\xi\,\bar\sigma^\mu\lambda\,+\,\bar\lambda\,\bar\sigma^\mu\xi),
\ers

Let us write (2) as ${\cal L}\equiv(1/4\pi)({\rm Im}\,\zeta)$. Furthermore, we write $\zeta=\zeta_1+\zeta_2$, where $\zeta_2$ contains the terms in $\lambda$ and/or $F^{\mu\nu}$ and $\zeta_1$ contains all the other terms.

We have
\brs
&&\delta(\pd\mu a\pu\mu\bar a-i\psi\sigma^\mu\pd\mu\bar\psi+f\bar f)\\
&&=\sqrt2\,\pd\mu\left[(\xi\psi)\pu\mu\bar a-2(\bar\psi\bar\sigma^{\mu\nu}\bar\xi)\pd\nu a+i\bar f(\bar\xi\bar\sigma^\mu\psi)\right]
\ers
so it is easy to calculate
$$
\delta\zeta_1=\sqrt2\,\pd\mu\left[\tau(\xi\psi)\pu\mu\bar a-2\bar\tau(\xi\sigma^{\mu\nu}\psi)\pd\nu\bar a-i\bar\tau f(\xi\sigma^\mu\bar\psi)\right]
$$
To evaluate $\delta\zeta_2$, the first step is to calculate
\brs
&&\delta\left[{1\over2}D^2-{1\over4}F_{\mu\nu}(F^{\mu\nu}-i\widetilde F^{\mu\nu})-i\lambda\sigma^\mu\pd\mu\bar\lambda\right]\\
&&=\pu\mu\left[(\bar\xi\bar\sigma^\nu\lambda)\left(\eta_{\mu\nu}D-i(F_{\mu\nu}-i\widetilde F_{\mu\nu})\right)\right]+2(\bar\xi\bar\sigma^\nu\lambda)(\pu\mu\widetilde F_{\mu\nu})
\ers
When the above variation is added to the other ingredients of $\delta\zeta_2$, namely $\delta[i\sqrt2D(\psi\lambda)-\sqrt2(\lambda\sigma^{\mu\nu}\psi)F_{\mu\nu}-(\lambda\lambda)f]$ and $\delta[(\psi\psi)(\lambda\lambda)]$, an expression containing 28 terms is obtained. Following a lengthy calculation whose full details will be given elsewhere [thesis], we obtain
\brs
\delta\zeta_2&=&\pd\mu\left[\tau D(\bar\xi\bar\sigma^\mu\lambda)-i\tau(\bar\xi\bar\sigma_\nu\lambda)(F^{\mu\nu}-i\widetilde F^{\mu\nu})\right.\\
&&\quad\left.+i\sqrt2\tau'(\psi\lambda)(\bar\xi\bar\sigma^\mu\lambda)\right]+2\tau(\bar\xi\bar\sigma^\nu\lambda)(\pu\mu\widetilde F_{\mu\nu})
\ers
Finally, from $\delta\zeta_1$ and $\delta\zeta_2$ we obtain $\delta{\cal L}$ as given by (\ref{deltaL}).

After obtaining the supercurrent that generates this variation from Noether's theorem, we construct the second \none supercurrent by using the $SU(2)_R$ symmetry of the \nn  vector multiplet. Under $SU(2)_R$, the parameters $\xi^{(1)}$ and $\xi^{(2)}$ and the fermion fields $\lambda$ and $\psi$ transform as doublets
$$
\xi^{(1)}\to\xi^{(2)},\qquad\xi^{(2)}\to-\xi^{(1)},\qquad \lambda\to\psi,\qquad \psi\to-\lambda,
$$
while $a$ and $A^\mu$ are singlets. These transformations induce the following transformations of the auxiliary fields:
$$
f\to\bar f,\qquad D\to-D.
$$
Alternatively, $SU(2)_R$ transformations could be applied to the variations of component fields and these could be used to construct the second current directly. We list these variations here because they are used in our derivations of BPS equations. Considering that in the field variations given at the beginning of this subsection, $\delta$ and $\xi$ refer to the first set of \none supersymmetry transformations and therefore should have a superscript (1), the second set of variations is
\brs
\delta^{(2)} a&=&-\sqrt{2}\,\xi^{(2)}\lambda\quad({\rm so~}\delta^{(2)}\tau=-\sqrt{2}\,\tau'\,\xi^{(2)}\lambda),\\
\delta^{(2)}\psi&=&-\sigma^{\mu\nu}\xi^{(2)} F_{\mu\nu}\,-\,i\xi^{(2)} D,\\
\delta^{(2)}\lambda&=&-i\sqrt{2}\,\sigma^\mu\,\bar\xi^{(2)}\,\partial_\mu a\,-\,\sqrt{2}\,\xi^{(2)}\bar f,\\
\delta^{(2)} A^\mu&=&i(\bar\xi^{(2)}\,\bar\sigma^\mu\psi\,-\,\bar\psi\,\bar\sigma^\mu\xi^{(2)}),\quad{\rm or}\\
\delta^{(2)} F^{\mu\nu}&=&i[(\xi^{(2)}\,\sigma^\nu \partial^\mu\bar\psi\,+\,\bar\xi^{(2)}\,\bar\sigma^\nu\partial^\mu\psi)\,-\,(\nu<\!\!\!--\!\!\!>\mu)].
\ers
Then, the \nn  supersymmetry variation of the component fields is obtained by adding the two \none variations; for example
$$
\delta a=\delta^{(1)} a+\delta^{(2)} a=\sqrt{2}\,(\xi^{(1)}\psi-\xi^{(2)}\lambda),
$$
etc.

\bigskip
\subsection{Variations of Component Fields from Supersymmetry Generators}
Using the supersymmetry generator (\ref{Q1}), we now reproduce the \none supersymmetry variations of the fields of the \nn  vector multiplet. (The second set of \none variations, generated by (\ref{Q2}), is obtained from the first one by a $SU(2)_R$ transformation, so it does not require a separate verification.) In the following, all anticommutators and commutators are Dirac brackets.

$$
\delta^{(1)}a=i\xi^{(1)\alpha}\{Q_\alpha^{(1)},a\}+i\bar\xi^{(1)}_{\dot\alpha}\{Q^{(1)\dot\alpha},a\}=\int\,d^3x\,i\xi^{(1)\alpha}\sqrt2[\Pi_a,a]\xi\psi=\sqrt2\xi\psi
$$
\brs
\delta^{(1)}\psi_\gamma&=&i\xi^{(1)\alpha}\{Q_\alpha^{(1)},\psi_\gamma\}+i\bar\xi^{(1)}_{\dot\alpha}\{Q^{(1)\dot\alpha},\psi_\gamma\}\\
&=&\int\,d^3x\,\left[i\xi^{(1)\alpha}\sqrt2\{\Pi_a\psi_\alpha,\psi_\gamma\}+{1\over4\pi}\,{i\over\sqrt2}\,\xi^{(1)\alpha}\bar\tau'(\sigma^0)_{\alpha\dot\alpha}\{(\bar\psi\bar\lambda)\bar\lambda^{\dot\alpha},\psi_\gamma\}\right.\\
&&\hspace{2cm}\left.+{1\over4\pi}\,i\sqrt2\,\bar\xi^{(1)}_{\dot\alpha}({\rm Im}\,\tau)(\bar\sigma^\mu\sigma^0)^{\dot\alpha}\,_{\dot\beta}\{\bar\psi^{\dot\beta},\psi_\gamma\}\pd\mu a\right]\\
&=&i\sqrt2(\sigma^\mu\bar\xi^{(1)})_\alpha\pd\mu a+\sqrt2\xi^{(1)}_\alpha\left(-{i\over4}\,{\tau'\over{\rm Im}\,\tau}\,\psi\psi\right)+{i\over\sqrt2}\,{\bar\tau'\over{\rm Im}\,\tau}(\xi^{(1)}\sigma^0\bar\lambda)(\bar\lambda\bar\sigma^0)^\delta\varepsilon_{\delta\alpha}\\
&=&i\sqrt2(\sigma^\mu\bar\xi^{(1)})_\alpha\pd\mu a+\sqrt2\xi^{(1)}_\alpha\left[-{i\over4{\rm Im}\,\tau}\,\left(\tau'\psi\psi-\bar\tau'\bar\lambda\bar\lambda\right)\right]\\
&=&i\sqrt2(\sigma^\mu\bar\xi^{(1)})_\alpha\pd\mu a+\sqrt2\xi^{(1)}_\alpha f
\ers
where we have used the expression of the auxiliary field $f$ given in section 2.
\brs
&&\\
\delta^{(1)}\lambda_\gamma&=&i\xi^{(1)\alpha}\{Q_\alpha^{(1)},\lambda_\gamma\}+i\bar\xi^{(1)}_{\dot\alpha}\{Q^{(1)\dot\alpha},\lambda_\gamma\}\\
&=&\int\,d^3x\,i\xi^{(1)\alpha}\left[\sqrt2\{\Pi_a\psi_\alpha,\lambda_\gamma\}-\sqrt2\,{\bar\tau'\over\bar\tau}\,\{(\bar\psi\bar\lambda)(\Pi_\lambda)_{\alpha},\lambda_\gamma\}\right.\\
&&\hspace{2cm}+\left.i\left(\Pi_{A_i}-{\bar\tau\over4\pi}\,\widetilde F^{0i}\right)(\sigma_i)_{\alpha\dot\alpha}\{\bar\alpha^{\dot\alpha},\lambda_\gamma\}\right]\\
&=&{i\over\sqrt2}\,{\tau'\over{\rm Im}\,\tau}\,(\xi^{(1)}\psi)\lambda_\gamma-{i\over\sqrt2}\,{\bar\tau'\over{\rm Im}\,\tau}\,\xi^{(1)}_\gamma(\bar\psi\bar\lambda)\\
&&+{i\over\sqrt2}\,{\bar\tau'\over{\rm Im}\,\tau}\,(\xi^{(1)}\sigma^0\bar\lambda)(\bar\psi\bar\sigma^0)^\delta\varepsilon_{\delta\gamma}-{4\pi\over{\rm Im}\,\tau}\,\left(\Pi_{A_i}-{\bar\tau\over4\pi}\,\widetilde F^{0i}\right)(\xi^{(1)}\sigma_i\bar\sigma^0)^\beta\varepsilon_{\beta\gamma}\\
&=&(F^{0i}-i\widetilde F^{0i})(\xi^{(1)}\sigma_i\bar\sigma^0)^\beta\varepsilon_{\beta\gamma}\\
&&+{i\over\sqrt2}\,{1\over{\rm Im}\,\tau}\,\left[(-\tau'\lambda\sigma^{0i}\psi+\bar\tau'\bar\lambda\bar\sigma^{0i}\bar\psi)(\xi^{(1)}\sigma_i\bar\sigma^0)^\beta\varepsilon_{\beta\gamma}\right.\\
&&+\left.\tau'(\xi^{(1)}\psi)\lambda_\gamma-\bar\tau'\xi^{(1)}_\gamma(\bar\psi\bar\lambda)+\bar\tau'(\xi^{(1)}\sigma^0\bar\lambda)(\bar\psi\bar\sigma^0)^\delta\varepsilon_{\delta\gamma}\right]
\ers
It is convenient to contract the square parenthesis with an arbitrary spinor $\theta^\gamma$, use Fierz identities and in the end extract the parenthesis from the result. Using the identities
$$
(\lambda\sigma^{0i}\psi)(\xi^{(1)}\sigma_i\bar\sigma^0\theta)=-(\xi^{(1)}\psi)(\lambda\theta)-{1\over2}(\psi\lambda)(\xi^{(1)}\theta),
$$$$
(\bar\lambda\bar\sigma^{0i}\bar\psi)(\xi^{(1)}\sigma_i\bar\sigma^0\theta)=-(\xi^{(1)}\sigma^0\bar\psi)(\theta\sigma^0\bar\lambda)+{1\over2}(\bar\psi\bar\lambda)(\xi^{(1)}\theta),
$$$$
(\xi^{(1)}\sigma^0\bar\lambda)(\theta\sigma^0\bar\psi)=-(\xi^{(1)}\sigma^0\bar\psi)(\theta\sigma^0\bar\lambda)+(\bar\psi\bar\lambda)(\xi^{(1)}\theta),
$$
and
$$
\sigma^{\mu\nu}F_ {\mu\nu}=-(\sigma^i\bar\sigma^0)(F_{0i}-i\widetilde F_{0i})
$$
we obtain
\brs
\delta^{(1)}\lambda_\gamma&=&-(\sigma^{\mu\nu})_\gamma\,^\alpha\bar\xi^{(1)}_\alpha F_{\mu\nu}+i\xi^{(1)}_\gamma\left[-{1\over2\sqrt{2}}\,{1\over{\rm Im}\,\tau}\left(\tau'\,\psi\lambda+\bar\tau'\,\bar\psi\bar\lambda\right)\right]\\
&=&-\frac{1}{2}\,\sigma^\mu\bar\sigma^\nu\xi F_{\mu\nu}\,+\,i\xi D,
\ers
where we have used the expression of the auxiliary field $D$ given in section 2. Finally,
\brs
\delta^{(1)}A^\mu&=&i\xi^{(1)\alpha}\{Q_\alpha^{(1)},A^\mu\}+i\bar\xi^{(1)}_{\dot\alpha}\{Q^{(1)\dot\alpha},A^\mu\}\\
&=&-\int\,d^3x\,\left(\left[\Pi_{A_\nu},A^\mu\right](\xi^{(1)}\sigma_\nu\bar\lambda+\bar\xi^{(1)}\bar\sigma_\nu\lambda)\right)\\
&=&i(\bar\xi^{(1)}\bar\sigma^\mu\lambda-\bar\lambda\bar\sigma^\mu\xi^{(1)})
\ers
so indeed we can recover the \nn  field variations using the supersymmetry generators.

\bigskip
\section{Calculation of the Central Charge in the U(1) effective theory}
\begin{eqnarray}
\left\{Q^{(1)}_\alpha({\bf x}),Q^{(2)}_\beta({\bf y})\right\}&=&\int\,d^3x\,\int\,d^3y\,\left[-2\left\{\Pi_a\psi_\alpha,\Pi_a\lambda_\beta\right\}\right.\label{a1}\\
&&-\frac{2}{4\pi}(\sigma^i\bar\sigma^0)_\beta\,^\gamma\partial_i\bar a\left\{\Pi_a\psi_\alpha,({\rm Im}\,\tau)\lambda_\gamma\right\}\label{a2}\\
&&+i\sqrt{2}\Pi_a\left(\Pi_{A_i}-\frac{\bar\tau}{4\pi}\widetilde F^{0i}\right)(\sigma_i)_{\beta\dot\beta}\left\{\psi_\alpha,\bar\psi^{\dot\beta}\right\}\label{a3}\\
&&+2\Pi_a\frac{\bar\tau'}{\bar\tau}\left\{\psi_\alpha,(\bar\psi\bar\lambda)(\Pi_\psi)_\beta\right\}\label{a4}\\
&&-\frac{2}{4\pi}(\sigma^i\bar\sigma^0)_\alpha\,^\delta\partial_i \bar a\left\{({\rm Im}\,\tau)\psi_\delta,\Pi_a\lambda_\beta\right\}\label{a5}\\
&&+\frac{i\sqrt{2}}{4\pi}({\rm Im}\,\tau)(\sigma^i\bar\sigma^0)_\alpha\,^\delta\partial_i\bar a\left(\Pi_{A_j}-\frac{\bar\tau}{4\pi}\widetilde F^{0j}\right)\nonumber\\
&&\qquad \times(\sigma_j)_{\beta\dot\beta}\left\{\psi_\delta,\bar\psi^{\dot\beta}\right\}\label{a6}\\
&&+\frac{2}{4\pi}({\rm Im}\,\tau)\frac{\bar\tau'}{\bar\tau}(\partial_i\bar a)(\sigma^i\bar\sigma^0)_\alpha\,^\delta\left\{\psi_\delta,(\bar\psi\bar\lambda)(\Pi_\psi)_\beta\right\}\label{a7}\\
&&-i\sqrt{2}\Pi_a\left(\Pi_{A_i}-\frac{\bar\tau}{4\pi}\widetilde F^{0i}\right)(\sigma_i)_{\alpha\dot\alpha}\left\{\bar\lambda^{\dot\alpha},\lambda_\beta\right\}\label{a8}\\
&&-\frac{i\sqrt{2}}{4\pi}({\rm Im}\,\tau)(\partial_j\bar a)\left(\Pi_{A_i} -\frac{\bar\tau}{4\pi}\widetilde F^{0i}\right)\nonumber\\
&&\qquad\times(\sigma_i)_{\alpha\dot\alpha}(\sigma^j\bar\sigma^0)_\beta\,^\gamma\left\{\bar\lambda^{\dot\alpha},\lambda_\gamma\right\}\label{a9}\\
&&+\frac{1}{4\pi}\bar\tau(\sigma_i)_{\alpha\dot\alpha}\bar\lambda^{\dot\alpha}(\sigma_j)_{\beta\dot\beta}\bar\psi^{\dot\beta}\,\left[\Pi_{A_i},\widetilde F^{0j}\right]\label{a10}\\
&&+\frac{1}{4\pi}\bar\tau(\sigma_i)_{\alpha\dot\alpha}\bar\lambda^{\dot\alpha}(\sigma_j)_{\beta\dot\beta}\bar\psi^{\dot\beta}\,\left[\widetilde F^{0i},\Pi_{A_j}\right]\label{a11}\\
&&+2\Pi_a\frac{\bar\tau'}{\bar\tau}\left\{(\bar\psi\bar\lambda)(\Pi_\lambda)_\alpha,\lambda_\beta\right\}\label{a12}\\
&&+\left.\frac{2}{4\pi}({\rm Im}\,\tau)\frac{\bar\tau'}{\bar\tau}(\partial_j\bar a)(\sigma^j\bar\sigma^0)_\beta\,^\gamma\left\{(\bar\psi\bar\lambda)(\Pi_\lambda)_\alpha,\lambda_\gamma\right\}\right]\label{a13}
\end{eqnarray}
The terms (\ref{a3}), (\ref{a6}), (\ref{a8}), and (\ref{a9}), involve directly the Dirac brackets given earlier in this appendix. We also have in (\ref{a1})
$$
\left\{\Pi_a\psi_\alpha,\Pi_a\lambda_\beta\right\}=-\delta^3({\bf x}-{\bf y})\frac{\tau'\Pi_a}{2\,{\rm Im}\,\tau}\left\{\psi_\alpha,\lambda_\beta\right\}=0
$$
In (\ref{a2}),
$$
\left\{\Pi_a\psi_\alpha,({\rm Im}\,\tau)\lambda_\gamma\right\}=-\delta^3({\bf x}-{\bf y})\frac{\tau'}{2}\left\{\psi_\alpha,\lambda_\gamma\right\}=0
$$
In (\ref{a5}),
$$
\left\{({\rm Im}\,\tau)\psi_\delta,\Pi_a\lambda_\beta\right\}=-\delta^3({\bf x}-{\bf y})\frac{\tau'}{2}\left\{\psi_\delta,\lambda_\beta\right\}=0
$$
In (\ref{a4}) and (\ref{a7}) we have
$$
\left\{\psi_\alpha,(\bar\psi\bar\lambda)(\Pi_\psi)_\beta\right\}=-\delta^{(3)}({\bf x}-{\bf y})\,\frac{4\pi}{{\rm Im}\,\tau}\,(\sigma^0)_{\alpha\dot\beta}\bar\lambda^{\dot\beta}(\Pi_\psi)_\beta+\delta^{(3)}({\bf x}-{\bf y})(\bar\psi\bar\lambda)\,\frac{\bar\tau}{2\,{\rm Im}\,\tau}\,\varepsilon_{\alpha\beta}
$$
In (\ref{a10}) and (\ref{a11}) it is useful to write in the spatial coordinates explicitly:
\begin{eqnarray*}
&\frac{1}{4\pi}\left\{\bar\tau({\bf y})\left(\sigma_i\bar\lambda({\bf x})\right)_\alpha\left(\sigma_j\bar\psi({\bf y})\right)_\beta\,\left[\Pi_{A_i}({\bf x}),\widetilde F^{0j}({\bf y})\right]\right.&\\
&+\left.\bar\tau({\bf x})\left(\sigma_i\bar\lambda({\bf x})\right)_\alpha\left(\sigma_j\bar\psi({\bf y})\right)_\beta\,\left[\widetilde F^{0i}({\bf x}),\Pi_{A_j}({\bf y})\right]\right\}&
\end{eqnarray*}
For the Lagrangian (2), the Dirac brackets for the gauge field are equal to the canonical brackets:
$$
\left[\Pi_{A_i}({\bf x}),\widetilde F^{0j}({\bf y})\right]=\varepsilon^{0jkl}\partial_{y^k}\left[\Pi_{A_i}({\bf x}),A_l({\bf y})\right]=i\varepsilon^{0ijk}\partial_{y^k}\delta^3({\bf x}-{\bf y})
$$
$$
\left[\widetilde F^{0i}({\bf x}),\Pi_{A_j}({\bf y})\right]=i\varepsilon^{0ijk}\partial_{x^k}\delta^3({\bf x}-{\bf y})
$$
Integrating (\ref{a10}) by parts and dropping fermionic boundary terms (which vanish because the fermion fields vary with distance as $\left|{\bf x}\right|^{-3/2}$), we obtain
\begin{eqnarray*}
&&\frac{i\varepsilon^{0ijk}}{4\pi}\int\,d^3x\,\left(\sigma_i\bar\lambda({\bf x})\right)_\alpha\int\,d\,^3y\,\bar\tau({\bf y})\left(\sigma_j\bar\psi({\bf y})\right)_\beta\,\partial_{y^k}\delta^3({\bf x}-{\bf y})\\
&&=\frac{i\varepsilon^{0ijk}}{4\pi}\int\,d^3x\,\left[\partial_{x^k}\left(\sigma_i\bar\lambda({\bf x})\right)_\alpha\right]\bar\tau({\bf x})\left(\sigma_j\bar\psi({\bf x})\right)_\beta
\end{eqnarray*}
Similarly, integrating (\ref{a11}) by parts,
\begin{eqnarray*}
&&\frac{i\varepsilon^{0ijk}}{4\pi}\int\,d^3x\,\bar\tau({\bf x})\left(\sigma_i\bar\lambda({\bf x})\right)_\alpha\,\left[\partial_{x^k}\delta^3({\bf x}-{\bf y})\right]\int\,d^3y\,\left(\sigma_j\bar\psi({\bf y})\right)_\beta\\
&&=\frac{i\varepsilon^{0ijk}}{4\pi}\int\,d^3x\,\bar\tau({\bf x})\left(\sigma_i\bar\lambda({\bf x})\right)_\alpha\partial_{x^k}\left(\sigma_j\bar\psi({\bf x})\right)_\beta
\end{eqnarray*}
From here on it is no longer necessary to write in explicitly the dependence on ${\bf x}$. Adding the results from (\ref{a10}) and (\ref{a11}), we obtain
$$
\frac{i\varepsilon^{0ijk}}{4\pi}\int\,d^3x\,\bar\tau\partial_{k}\left((\sigma_i\bar\lambda)_\alpha(\sigma_j\bar\psi)_\beta\right)=-\frac{i\varepsilon^{0ijk}}{4\pi}\int\,d^3x\,(\partial_{k}\bar\tau)(\sigma_i\bar\lambda)_\alpha(\sigma_j\bar\psi)_\beta
$$
where in the last step we have again dropped a fermionic boundary term. Using the identities
\beq\label{id}
(\sigma^\mu)_{\alpha\dot\alpha}(\sigma^\nu)_{\beta\dot\beta}-(\sigma^\nu)_{\alpha\dot\alpha}(\sigma^\mu)_{\beta\dot\beta}=2\left[(\sigma^{\mu\nu}\varepsilon)_{\alpha\beta}\varepsilon_{\dot\alpha\dot\beta}+(\varepsilon\bar\sigma^{\mu\nu})_{\dot\alpha\dot\beta}\varepsilon_{\alpha\beta}\right]
\eeq
and
$$
\varepsilon^{\mu\nu\rho\lambda}\sigma_{\rho\lambda}=-2i\sigma^{\mu\nu},\quad\varepsilon^{\mu\nu\rho\lambda}\bar\sigma_{\rho\lambda}=2i\bar\sigma^{\mu\nu},
$$
we can further write the sum of (\ref{a10}) and (\ref{a11}) as
\begin{eqnarray*}
&&-\frac{i}{4\pi}\,\frac{1}{2}\,\varepsilon^{0ijk}\int\,d^3x\,(\partial_{k}\bar\tau)\left[(\sigma_i\bar\lambda)_\alpha(\sigma_j\bar\psi)_\beta-(\sigma_j\bar\lambda)_\alpha(\sigma_i\bar\psi)_\beta\right]\\
&&=-\frac{i}{8\pi}\,\varepsilon^{0ijk}\int\,d^3x\,(\partial_{k}\bar\tau)\left[(\sigma_i)_{\alpha\dot\alpha}(\sigma_j)_{\beta\dot\beta}-(\sigma_j)_{\alpha\dot\alpha}(\sigma_i)_{\beta\dot\beta}\right]\bar\lambda^{\dot\alpha}\bar\psi^{\dot\beta}\\
&&=-\frac{i}{4\pi}\int\,d^3x\,(\partial_{k}\bar\tau)\left[\varepsilon^{0ijk}(\sigma_{ij})_\alpha\,^\gamma\varepsilon_{\gamma\beta}\varepsilon_{\dot\alpha\dot\beta}+\varepsilon_{\dot\alpha\dot\gamma}\varepsilon^{0ijk}(\bar\sigma_{ij})^{\dot\gamma}\,_{\dot\beta}\varepsilon_{\alpha\beta}\right]\bar\lambda^{\dot\alpha}\bar\psi^{\dot\beta}\\
&&=-\frac{i}{4\pi}\int\,d^3x\,(\partial_{k}\bar\tau)\left[-2i(\sigma^{0k})_\alpha\,^\gamma\varepsilon_{\gamma\beta}\varepsilon_{\dot\alpha\dot\beta}+2i\varepsilon_{\dot\alpha\dot\gamma}(\bar\sigma^{0k})^{\dot\gamma}\,_{\dot\beta}\varepsilon_{\alpha\beta}\right]\bar\lambda^{\dot\alpha}\bar\psi^{\dot\beta}\\
&&=\frac{2}{4\pi}\int\,d^3x\,(\partial_{k}\bar\tau)\left[(\bar\psi\bar\lambda)(\sigma^{0k})_\alpha\,^\gamma\varepsilon_{\gamma\beta}-(\bar\lambda\bar\sigma^{0k}\bar\psi)\varepsilon_{\alpha\beta}\right]
\end{eqnarray*}
Finally, in (\ref{a12}) and (\ref{a13}) we have
$$
\left\{(\bar\psi\bar\lambda)(\Pi_\lambda)_\alpha,\lambda_\beta\right\}=-\delta^{(3)}({\bf x}-{\bf y})\,\frac{4\pi}{{\rm Im}\,\tau}\,\bar\psi_{\dot\beta}(\bar\sigma^0)^{\dot\beta\delta}\varepsilon_{\delta\beta}(\Pi_\lambda)_\alpha-\delta^{(3)}({\bf x}-{\bf y})(\bar\psi\bar\lambda)\,\frac{\bar\tau}{2\,{\rm Im}\,\tau}\,\varepsilon_{\alpha\beta}
$$
Collecting all these results, we end up with
\begin{eqnarray}
\left\{Q^{(1)}_\alpha,Q^{(2)}_\beta\right\}&=&\int\,d^3x\,\left[i\sqrt{2}(\sigma^i\bar\sigma^0)_\alpha\,^\delta(\partial_i\bar a)\left(\Pi_{A_j}-\frac{\bar\tau}{4\pi}\widetilde F^{0j}\right)(\sigma_j\bar\sigma^0)_\beta\,^\gamma\varepsilon_{\gamma\delta}\right.\label{a14}\\
&&-2\,\frac{\bar\tau'}{\bar\tau}\,(\partial_i\bar a)(\sigma^i\bar\sigma^0)_\alpha\,^\delta(\sigma^0\bar\lambda)_\delta(\Pi_\psi)_\beta\label{a15}\\
&&+\frac{\bar\tau'}{4\pi}\,(\partial_i\bar a)(\sigma^i\bar\sigma^0)_\alpha\,^\delta(\bar\psi\bar\lambda)\varepsilon_{\delta\beta}\label{a16}\\
&&-i\sqrt{2}(\sigma^j\bar\sigma^0)_\beta\,^\delta(\partial_j\bar a)\left(\Pi_{A_i}-\frac{\bar\tau}{4\pi}\widetilde F^{0i}\right)(\sigma_i\bar\sigma^0)_\alpha\,^\gamma\varepsilon_{\gamma\delta}\label{a17}\\
&&+\frac{2}{4\pi}\,(\partial_{k}\bar\tau)\left[(\bar\psi\bar\lambda)(\sigma^{0k})_\alpha\,^\gamma\varepsilon_{\gamma\beta}-(\bar\lambda\bar\sigma^{0k}\bar\psi)\varepsilon_{\alpha\beta}\right]\label{a18}\\
&&+\frac{\bar\tau'}{4\pi}\,(\partial_j\bar a)(\sigma^j\bar\sigma^0)_\beta\,^\delta(\bar\psi\bar\lambda)\varepsilon_{\delta\alpha}\label{a19}\\
&&\left.-2\,\frac{\bar\tau'}{\bar\tau}\,(\partial_j\bar a)(\sigma^j\bar\sigma^0)_\beta\,^\gamma(\bar\psi\bar\sigma^0)^\delta\varepsilon_{\delta\gamma}(\Pi_\lambda)_\alpha\right]\label{a20}
\end{eqnarray}
The terms (\ref{a16}) and (\ref{a19}) add up to
\begin{equation}\label{a21}
-\frac{4}{4\pi}(\partial_{i}\bar\tau)(\bar\psi\bar\lambda)(\sigma^{0i})_\alpha\,^\gamma\varepsilon_{\gamma\beta}
\end{equation}
The terms (\ref{a15}) and (\ref{a20}) add up to
\begin{equation}\label{a22}
\frac{1}{4\pi}(\partial_{i}\bar\tau)\left[(\sigma^i\bar\lambda)_\alpha(\sigma^0\bar\psi)_\beta+(\sigma^i\bar\psi)_\beta(\sigma^0\bar\lambda)_\alpha\right]
\end{equation}
Applying again the identity (\ref{id}), we can combine (\ref{a21}) and (\ref{a22}) to obtain
$$
\frac{2}{4\pi}(\partial_{i}\bar\tau)\left[(\bar\lambda\bar\sigma^{0i}\bar\psi)\varepsilon_{\alpha\beta}-(\bar\psi\bar\lambda)(\sigma^{0i})_\alpha\,^\gamma\varepsilon_{\gamma\beta}\right],
$$
which cancels (\ref{a18}). The only surviving terms are (\ref{a14}) and (\ref{a17}) which give
$$
\left\{Q^{(1)}_\alpha,Q^{(2)}_\beta\right\}=-i(2\sqrt{2})\int\,d^3x\,(\partial_i\bar a)\left(\Pi_{A_i}-\frac{\bar\tau}{4\pi}\widetilde F^{0i}\right)\varepsilon_{\alpha\beta}
$$
Comparison with the \nn  supersymmetry algebra, eq. (\ref{alg2}), gives the following central charge formula
$$
Z=\int\,d^3x\,(\partial_i a)\left(\Pi_{A_i}-\frac{\tau}{4\pi}\,\widetilde F^{0i}\right)
$$
In section 5 we have explained how this expression finally reduces to
$$
Z\,=\,a_\infty\,n_e\,+\,a_{D\infty}\,n_m.
$$

\bigskip
\section{Modular Transformations and Classical Duality Rotations}
We derive the relation between the rotation angle $\varphi$ of a classical duality transformation (the SO(2) duality) and the parameters of a modular transformation (the $PSL(2,\Z)$ duality, at the quantum level).

The classical electric-magnetic duality is an SO(2) rotation of the electric and magnetic charges and fields. (In this section, a prime will refer to duality-transformed quantities.) The Maxwell equations with electric and magnetic sources are invariant under a simultaneous rotation of fields and charges:
$$
{\bf E}\,'+i\,{\bf B}\,'=e^{-i\varphi}({\bf E}+i\,{\bf B})
$$
$$
Q_e\,'+i\,Q_m\,'=e^{-i\varphi}(Q_e+i\,Q_m)
$$
(The sign convention is such that ${\bf E}\to{\bf B}$ and ${\bf B}\to-{\bf E}$ for $\varphi=\pi/2$.) As it is well known, the Dirac-Schwinger-Zwanziger quantization condition, written as
$$
Q_e+i\,Q_m=\sqrt{\frac{4\pi}{{\rm Im}\,\tau}}\,(n_e+\tau\,n_m),
$$
where
$$
\tau=\frac{\theta}{2\pi}+i\,\frac{4\pi}{g^2},
$$
implies that states with charges $(Q_e,Q_m)$ are placed on a two dimensional lattice with periods $e$ and $\tau e$, and are represented by the vector $(n_e,n_m)$. This lattice breaks the classical duality symmetry SO(2) and can be described in terms of different fundamental cells. Choosing a different cell amounts to transforming $(n_e,n_m)$ by a matrix
$$
D=\left(\begin{array}{cc}
\alpha&\beta\\
\gamma&\delta
\end{array}\right)\;\in\;PSL(2,\Z)
$$
where $PSL(2,\Z)=SL(2,\Z)/\Z_2$ is the group of all analytic automorphisms of the upper half complex plane, called the modular group.
The transformation is usually given in terms of the inverse matrix,
$$
D^{-1}=\left(\begin{array}{cc}
\delta&-\beta\\
-\gamma&\alpha
\end{array}\right),
$$
as follows
$$
(n_m',n_e')=(n_m,n_e)D^{-1}=(n_m\delta\,-\,n_e\gamma\,,\,n_e\alpha\,-\,n_m\beta)
$$

The action of $PSL(2,\Z)$ on the charge lattice is implemented by modular transformations of $\tau$,
$$
\tau\to\frac{\alpha\tau+\beta}{\gamma\tau+\delta}\quad{\rm where~} \left(\begin{array}{cc}
\alpha&\beta\\
\gamma&\delta
\end{array}\right)\;\in\;PSL(2,\Z)
$$
In order to relate the matrix elements of $D\in PSL(2,\Z)$ to the rotation angle $\varphi$ of the classical duality group SO(2), let us consider the transformation of $n_e+\tau\,n_m$,
$$
n_e+\tau\,n_m\to n_e'+\frac{\alpha\tau+\beta}{\gamma\tau+\delta}\,n_m'=\frac{n_e+\tau\,n_m}{\gamma\tau+\delta}
$$
Then
$$
Q_e+i\,Q_m\to Q_e\,'+i\,Q_m\,'=\sqrt{\frac{4\pi}{{\rm Im}\,\tau'}}\,(n_e'+\tau'n_m')
$$
Note that the upper half complex plane is stable under the action of the modular group, so ${\rm Im}\,\tau'$ is always positive,
$$
{\rm Im}\,\tau'=\frac{{\rm Im}\,\tau}{\left|\gamma\tau+\delta\right|^2}
$$
Then
\brs
Q_e\,'+i\,Q_m\,'&=&\sqrt{\frac{4\pi}{{\rm Im}\,\tau}}\,\left|\gamma\tau+\delta\right|\,\frac{n_e+\tau\,n_m}{\gamma\tau+\delta}\\
&=&\frac{\left|\gamma\tau+\delta\right|}{\gamma\tau+\delta}(Q_e+i\,Q_m)
\ers
Comparison with $Q_e\,'+i\,Q_m\,'=e^{-i\varphi}(Q_e+i\,Q_m)$ shows that $\varphi$ {\em is the phase of} $\gamma\tau+\delta$,
\beq\label{angle}
e^{i\varphi}\,=\,\frac{\gamma\tau+\delta}{\left|\gamma\tau+\delta\right|}
\eeq
In other words, a modular transformation rotates the electric and magnetic fields and charges by
$$
\varphi={\rm arg}\,(\gamma\tau+\delta)={\rm arc\,tan}\,\frac{\gamma\,{\rm Im}\,\tau}{\gamma\,Re\,\tau+\delta}
$$
The scalar field $a$ and its dual $a_D$ transform as a doublet under $PSL(2,\Z)$,
$$
\left(\begin{array}{c}
a_D\,'\\a\,'
\end{array}\right)=D\,\left(\begin{array}{c}
a_D\\
a
\end{array}\right)
$$
so the central charge is duality-invariant,
$$
Z'=a\,'n_e'\,+\,a_D\,'n_m'=(n_m,n_e)\,D^{-1}\,D\,\left(\begin{array}{c}
a_D\\a\end{array}\right)=a\,n_e\,+\,a_D\,n_m=Z
$$
Since $M=\sqrt2\left|Z\right|$, the mass of BPS state is also duality-invariant. Another way to see this is by using (\ref{eandm}),
$$
M=\int\,d^3 x\,(E^2+B^2),
$$
in which the integrand is duality-invariant,
$$
E^2+B^2=({\bf E}+i{\bf B})\cdot({\bf E}-i{\bf B})\to e^{-i\varphi}({\bf E}+i\,{\bf B})\cdot e^{i\varphi}({\bf E}-i{\bf B})=E^2+B^2.
$$

The group $PSL(2,\Z)$ is generated by the action of two elements,
$$
T=\left(\begin{array}{cc}
1&1\\
0&1
\end{array}\right)\quad{\rm and~}\quad
S=\left(\begin{array}{cc}
0&-1\\
1&0
\end{array}\right)
$$

Under a T-transformation, we have
$$
\tau\to\tau+1,\quad {\rm Im}\,\tau\to {\rm Im}\,\tau,\quad (n_m,n_e)\to(n_m,n_e-n_m)
$$
and $\varphi=0$, so ${\bf E}\to{\bf E},\;{\bf B}\to{\bf B}$.

Under a S-transformation, we have
$$
\tau\to-{1\over\tau},\quad {\rm Im}\,\tau\to{{\rm Im}\,\tau\over\left|\tau\right|^2},\quad (n_m,n_e)\to(-n_e,n_m)
$$
and $\varphi={\rm arg}\,\tau$; only for $\tau$ purely imaginary ($\theta=0$), we have ${\bf E}\to{\bf B},\;{\bf B}\to-{\bf E}$.

In section 4 we have shown that the equation of motion (\ref{eqfora}) is not invariant under a general duality tranformation (for example, it is not S-invariant), but only under a subgroup of $PSL(2,\Z)$ consisting of the matrices
$$
\left(\begin{array}{cc}
1&\beta\\
0&1
\end{array}\right)
$$
The elements of a symmetry group of an equation transform solutions of the equation into other solutions. This symmetry group preserves $n_m$,
$$
(n_m,n_e)\to(n_m,n_e-n_m\beta)
$$
Since a magnetic monopole $(1,0)$ is a solution of the equations of motion, it follows that all dyons $(1,-\beta)$ are also solutions, with the same $a$ and with $a_D\to a_D+\beta a$. Therefore, given a monopole solution, we can construct a dyon solution of unit magnetic charge by T-duality.

\vspace{2cm}

\bigskip
\bibliographystyle{JHEP}

\begin{thebibliography}{999}

\bibitem{SW1}
N.~Seiberg and E.~Witten,
\emph{Electric-magnetic duality, monopole condensation, and
confinement in N=2 supersymmetric Yang-Mills theory},
Nucl. Phys. {\bf B426} (1994) 19, 
\hepth/{9407087}.

\bibitem{SW2}
N.~Seiberg and E.~Witten,
\emph{Monopoles, duality and chiral symmetry breaking in N=2 supersymmetric QCD}, Nucl. Phys. {\bf B431} (1994) 484,
\hepth/{9408099}.

\bibitem{Bergman}
O. Bergman,
\emph{Stable non-BPS dyons in N=2 SYM},
JHEP 9905 (1999) 004, \hepth/{9811064}.

\bibitem{CRV}
G. Chalmers, M. Ro\v cek and R. von Unge,
\emph{Monopoles in quantum corrected N=2 super Yang-Mills theory},
\hepth/{9612195}.

\bibitem{KLMVW}
A. Klemm, W. Lerche, P. Mayr, C. Vafa and N. Warner,
\emph{Selfdual strings and N=2 supersymmetric field theory},
\npb{477}{1996}{746},
\hepth{9604034}.

\bibitem{Sen}
A. Sen, 
\emph{F-theory and orientifolds},
\npb{475}{1996}{562},
\hepth{9605150}.

\bibitem{Faya} 
A. Fayyazuddin, 
\emph{Results in supersymmetric field theory from 3-brane probe
in F-theory},
\npb{497}{1997}{101},
\hepth{9701185}.

\bibitem{Bergman1} 
O. Bergman, 
\emph{Three-pronged strings and 1/4 BPS states in N=4 super-Yang-Mills 
theory},
\npb{525}{1998}{104},
\hepth{9712211}.

\bibitem{GHZ} 
M. Gaberdiel, T. Hauer and B. Zwiebach,
\emph{Open string--string junction transitions},
\npb{525}{1998}{117},
\hepth{9801205}.

\bibitem{BF1} 
O. Bergman and A. Fayyazuddin,
\emph{String junctions and BPS states in Seiberg-Witten theory},
\npb{531}{1998}{108},
[\hepth{9802033}].

\bibitem{MNS} 
A. Mikhailov, N. Nekrasov and S. Sethi, 
\emph{Geometric realization of BPS states in N=2 theories},
\npb{531}{1998}{345},
\hepth{9803142}.

\bibitem{DHIZ} 
O. DeWolfe, T. Hauer, A. Iqbal and B. Zwiebach,
\emph{Constraints on the BPS spectrum of N=2, D=4 theories with A-D-E flavor
symmetries},
\npb{534}{1998}{261},
\hepth{9805220}.

\bibitem{BF2} 
O. Bergman and A. Fayyazuddin,
\emph{String junction transitions in the moduli space of N=2 SYM},
\npb{535}{1998}{139},
\hepth{9806011}.

\bibitem{SV}
A. Shapere and C. Vafa,
\emph{The structure of Argyres-Douglas superconformal theories},
\hepth{9910182}.

\bibitem{RSVV}
A. Ritz, M. Shifman, A. Vainshtein and M. Voloshin,
\emph{Marginal stability and the metamorphosis of BPS states},
\hepth{0006028}.

\bibitem{ArgyresNarayan}
P.~C.~Argyres and K.~Narayan,
\emph{String webs from field theory},
\hepth/0101114.

\bibitem{RV}
A.~Ritz and A.~Vainshtein,
\emph{Long range forces and supersymmetric bound states},
hep-th/0102121.

\bibitem{Iorio2}
A. Iorio,
\emph{Supersymmetric Noether Currents and Seiberg-Witten Theory},
Ph. D. Thesis, \hepth/0006198. 

\bibitem{Iorio1}
A. Iorio,
\emph{A note on Seiberg-Witten central charge},
Phys.\ Lett.\ B {\bf 487} (2000) 171, 
\hepth/9905069.


\bibitem{GKMTZ}
J.~P.~Gauntlett, C.~Koehl, D.~Mateos, P.~K.~Townsend and M.~Zamaklar,
\emph{``Finite energy Dirac-Born-Infeld monopoles and string junctions},''
Phys.\ Rev.\ D {\bf 60} (1999)  045004, 
\hepth/9903156.

\bibitem{Denef}
F.~Denef,
\emph{Supergravity flows and D-brane stability},
JHEP{\bf 0008} (2000) 050,
\hepth/0005049.

\bibitem{DGR}
F.~Denef, B.~Greene and M.~Raugas,
\emph{Split attractor flows and the spectrum of BPS D-branes on the quintic},
\hepth/0101135.

\bibitem{Wei}
S. Weinberg,
\emph{The Quantum Theory of Fields}, Vol. 3:\emph{Supersymmetry},
Cambridge University Press, 1999.


\bibitem{Sri}
P. Srivastava
\emph{Supersymmetry, Superfields and Supergravity},
Adam Hilger, 1986.

\bibitem{HandT}
M.~Henneaux and C.~Teitelboim, 
\emph{Quantization of Gauge Systems},
Princeton University Press, 1992.

\bibitem{WE}
E.~Witten,
\emph{Dyons Of Charge E Theta / 2 Pi},
Phys.\ Lett.\ B {\bf 86}, 283 (1979).

\bibitem{Alv}
L. Alvarez-Gaum\'{e} and S. F. Hassan,
\emph{Introduction to S-Duality in N=2 Supersymmetric Gauge Theories},
Fortsch. Phys. {\bf 45} (1997) 159,
\hepth/{9701069}.

\bibitem{Wolf}
S. Wolf,
\emph{Computation of the Central Charge for the Leading Order of the N=2 Super-Yang-Mills Effective Action},
Mod.\ Phys.\ Lett.\ A {\bf 14} (1999)  2789,
\hepth/9905194.

\bibitem{WB}
J. Wess and J. Bagger,
\emph{Supersymmetry and Supergravity}, 2nd edition,
Princeton University Press, 1992.

\bibitem{BL}
D. Bailin and A. Love,
\emph{Supersymmetric Gauge Field Theory and String Theory},
IOP Publishing, 1994.

\end{thebibliography}

\end{document}